\documentclass[journal]{IEEEtran}
\usepackage{amsmath}
\usepackage{amssymb}
\usepackage{algorithm}
\usepackage{algorithmic}
\usepackage{booktabs}
\usepackage{capt-of}
\usepackage{graphicx}
\usepackage{multirow}
\usepackage{array}
\usepackage{xcolor}
\usepackage{hyperref}

\newcommand{\SR}{\mathrm{SR}}
\newcommand{\sg}{\mathrm{sg}}
\newcommand{\KL}{D_{\mathrm{KL}}}

\title{DSWM: Decomposed Spatio-Temporal World Models for Demand-Driven UAV Base Station Repositioning}
\author{Shengjie Zhong,~\IEEEmembership{Graduate Student Member,~IEEE,}
        Zhongliang Zhao,
        Jingxuan Chen,
        Xianbin Cao,
        Xinmei Qiang,
        Dapeng O. Wu,~\IEEEmembership{Fellow,~IEEE},
        Tony Q. S. Quek,~\IEEEmembership{Fellow,~IEEE}
\thanks{This work was supported by the Smart Grid-National Science and Technology Major Project (2026ZD0811400). (Corresponding author: Zhongliang Zhao.)}
\thanks{S. Zhong is with the School of Electronic and Information Engineering, Beihang University, Beijing 100191, China (e-mail: zhongshengjie@buaa.edu.cn).}
\thanks{Z. Zhao is with the School of Electronic and Information Engineering, Beihang University, Beijing, with the Department of Strategic and Advanced Interdisciplinary Research, Peng Cheng Laboratory, with the Hangzhou International Innovation Institute, Beihang University, Hangzhou, with the MIIT Key Lab. of Aerospace Mobile Communication, Beijing, and also with the State Key Lab. of CNS/ATM, Beijing (e-mail: zhaozl@buaa.edu.cn).}
\thanks{J. Chen is with the Department of Computer Science, City University of Hong Kong, Hong Kong, SAR, China, and also with the Hong Kong Generative AI Research and Development Center, Hong Kong, SAR, China (e-mail: chenjingxuan@buaa.edu.cn).}
\thanks{X. Cao is with the School of Electronic and Information Engineering, Beihang University, Beijing, and also with the MIIT Key Lab. of Aerospace Mobile Communication, Beijing.}
\thanks{X. M. Qiang is with the School of Electronic and Information Engineering, Beihang University, Beijing 100191, China (e-mail: qiangxm@buaa.edu.cn).}
\thanks{D. O. Wu is with the Department of Computer Science, The City University of Hong Kong, Hong Kong, SAR, China (e-mail: dpwu@ieee.org).}
\thanks{T. Q. S. Quek is with the Information Systems Technology and Design Pillar, Singapore University of Technology and Design, Singapore 487372 (e-mail: tonyquek@sutd.edu.sg).}
}

\begin{document}
\IEEEspecialpapernotice{This work has been submitted to the IEEE for possible publication. Copyright may be transferred without notice, after which this version may no longer be accessible.}
\maketitle

\begin{abstract}
Uncrewed aerial vehicle base stations (UAV-BSs) are expected to cover traffic demand that shifts across space and time, yet most repositioning schemes either re-solve an optimization problem per slot or learn reactive policies without an explicit demand model. We cast demand-driven fleet repositioning as latent-space decision-time planning and propose DSWM, \textbf{a decomposed spatio-temporal world model}: an agentic controller that perceives the demand field through a rolling observation window, retains operational context in a latent recurrent state, reasons about candidate motions by imagined rollouts under an uncertainty penalty, and coordinates the fleet through replanned first actions. 
DSWM learns a recurrent state-space model shaped by an exponential-moving-average (EMA) based latent predictive objective with variance regularization. It attaches a differentiable service simulator that replays the association, probabilistic line-of-sight channel, and Shannon rate chain inside latent rollouts. Planning uses a cross-entropy method whose imagined demand is anchored on the current observation window with mixing coefficient $\rho=0.95$. On a unified pipeline over three real datasets (Milan CDR (call detail record), Shanghai Telecom, YJMob100K) and 14 methods including five reproduced IEEE baselines, DSWM attains weekday served ratios of 0.889, 0.908, and 0.898, ranking first among non-ablated configurations on every dataset. On Milan it improves over the strongest non-learning baseline (Greedy, 0.780) by 0.109, a margin that comes from decision-time use of observations rather than prediction accuracy. 
\end{abstract}

\section{Introduction}
\IEEEPARstart{T}{he} low-altitude economy and 6G research agendas both position uncrewed aerial vehicle base stations (UAV-BSs) as a fast way to add capacity where ground infrastructure is absent or saturated \cite{mozaffari2019tutorial,ning2025communication}. UAV-BSs fly at low altitude, establish line-of-sight (LoS) links with high probability, and can be repositioned within minutes \cite{mozaffari2019tutorial}. Recent surveys argue that such aerial access layers must be driven by generative and predictive models of network state rather than static rules \cite{khoramnejad2025generative,ning2025communication}. The practical question is therefore not whether to move the fleet, but how to decide where each UAV-BS should be at every slot so that a demand map that drifts over hours is covered. This is an agentic control problem: the controller must perceive a partially observed demand field, retain operational context over hours, reason under uncertainty about delayed consequences, and coordinate a fleet under coupled energy and coverage constraints. Two properties make this decision hard. Demand is both spatially uneven and temporally nonstationary, with commute peaks, weekend shifts, and holiday anomalies. And each repositioning decision has delayed consequences: a move that takes minutes pays off only after the demand arrives, while battery limits force charging detours that remove capacity for an hour. Any controller must therefore trade current service against future service under dynamics it cannot observe directly.

Three lines of work address this question, and each leaves a specific gap. First, optimization-based designs decompose placement, trajectory, and resource allocation into subproblems solved by block coordinate descent (BCD) or convex relaxations \cite{zeng2017energy,alhourani2014optimal,sun2024multi,jia2025dynamic}. They are strong within a static or slowly changing demand snapshot, but they re-solve a nonconvex problem per decision and degrade when the demand drifts between solves. Second, model-free reinforcement learning (RL) and multi-agent RL (MARL) learn reactive policies for trajectory and resource control \cite{cui2019multi,qin2021distributed,wang2021deep,zhao2022multi,ning2023multi,feng2024graph,song2022evolutionary,wang2026lags,yuan2025hierarchical}. These policies carry no explicit demand model, so they need many environment samples and generalize weakly out of distribution (OOD). Third, two-stage pipelines first predict traffic with a spatio-temporal forecaster and then plan against the prediction \cite{yao2021mvstgn,gu2023spatial,wang2024uncertainty,liu2025stllm,zhang2024chattraffic}. Prediction error passes directly into the plan, and the forecasting objective (per-cell accuracy) is misaligned with the coverage objective (demand-weighted service).

\begin{figure}[!t]
\centering
\includegraphics[width=0.9\columnwidth]{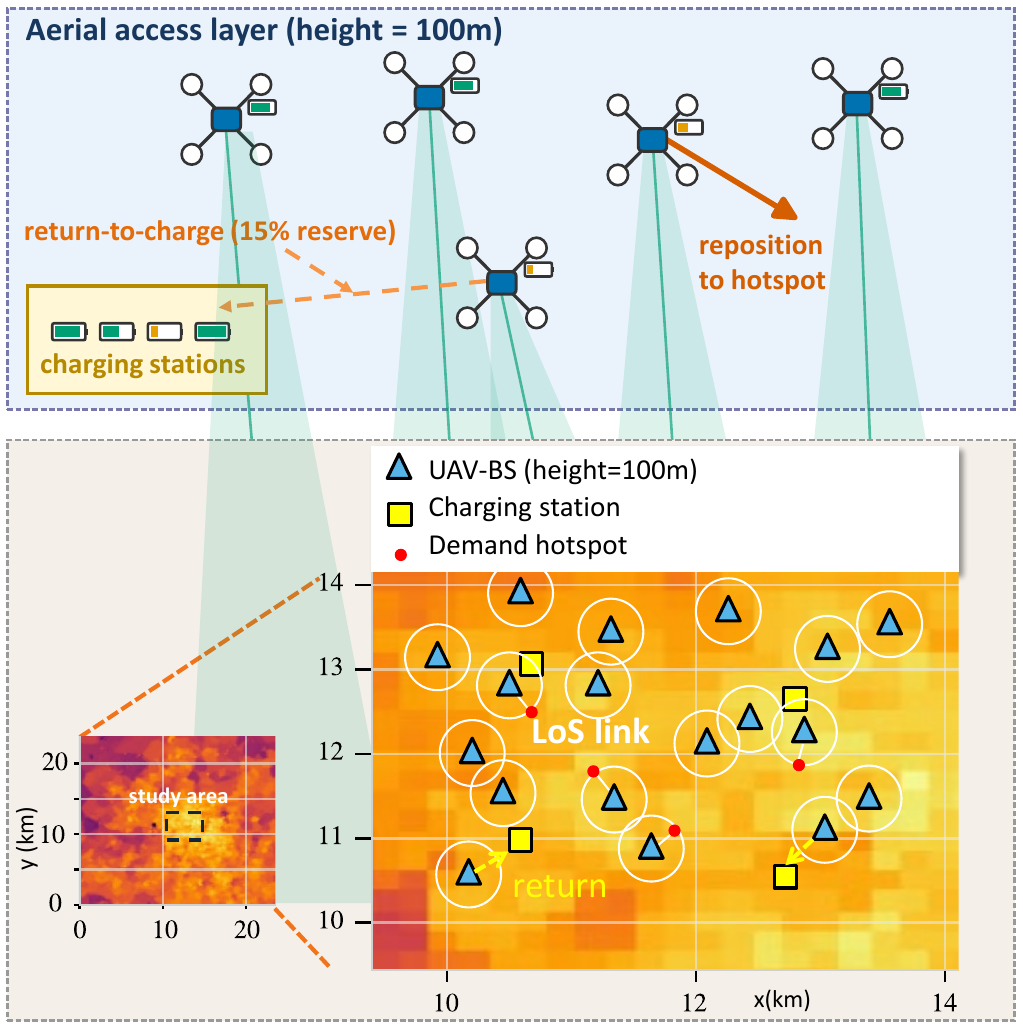}
\caption{Scenario and motivation. A fleet of $K$ UAV-BSs at altitude $h_{\mathrm{fly}}$ covers a gridded demand map $x_t$ that drifts over the day (commute bimodal pattern). Because the demand map is strongly structured in space and time, the bottleneck for coverage is how the current observation window is exploited at decision time, not raw forecasting accuracy. DSWM therefore plans in a learned latent space with the imagination anchored on the current observation.}
\label{fig:scenario}
\end{figure}

Our key insight, illustrated in Fig.~\ref{fig:scenario}, is that urban demand has strong spatio-temporal structure and is largely predictable, so the bottleneck is not prediction accuracy but how observations are used at decision time. World models answer exactly this question: they learn latent dynamics and let the agent plan by imagination at each step \cite{hafner2019learning,hansen2022temporal,hafner2025mastering}. If a planner anchors its imagined demand on the current observation window, it can in principle match a planner that sees the future. In a preliminary smaller-scale setting ($K{=}4$, without the battery model), anchoring improved both splits by $+0.018$ over the non-anchored variant at no cost (replicated from two checkpoints), while demand-forecasting enhancements were net negative; we report these as motivation---the main-table evidence appears in Section~\ref{sec:main}.
Based on this insight, we propose DSWM (Decomposed Spatio-Temporal World Model). DSWM learns a recurrent state-space model (RSSM) of the demand-service dynamics, shaped by an EMA-target latent predictive objective with a variance-margin regularizer (JEPA/VICReg-style) \cite{assran2025vjepa2} and variance regularization. A decomposed service head replays the environment's service physics---association, probabilistic LoS, Shannon rate, and capped per-cell fulfillment---as a differentiable simulator inside latent rollouts, and an ensemble disagreement term penalizes uncertain branches. A cross-entropy method (CEM) planner scores $H$-slot imagined trajectories with the demand anchored on the current observation with mixing coefficient $\rho=0.95$, and executes only the first action.

The contributions are summarized as follows:
\begin{itemize}
\item \textbf{Modeling paradigm (Section~\ref{sec:method}, Fig.~\ref{fig:framework}).} We formalize demand-driven UAV-BS fleet repositioning as latent-space decision-time planning and propose DSWM, an RSSM world model shaped by a latent predictive regularization objective. To our knowledge this is the first world-model formulation of UAV-BS repositioning with an explicit coverage objective.
\item \textbf{Architecture (Section~\ref{sec:decomp}, Fig.~\ref{fig:decomp}).} We design a decomposed demand-service physics head, a differentiable replica of the service simulator, with an ensemble uncertainty penalty. Ablation attributes $+0.120$ of served ratio to the decomposed head (0.769 vs.\ 0.889) and $-0.004$ to the uncertainty term, which we report as an insurance mechanism rather than a gain source.
\item \textbf{Algorithm (Section~\ref{sec:planning}, Fig.~\ref{fig:anchor}).} We propose observation-anchored CEM decision-time planning with anchor mixing coefficient $\rho=0.95$. DSWM improves over the strongest non-learning baseline by 0.109 on Milan (0.889 vs.\ 0.780); in a preliminary smaller-scale setting ($K{=}4$, without the battery model), anchoring improved both splits by $+0.018$ over the non-anchored variant at no cost (replicated from two checkpoints).
\item \textbf{Benchmark evidence (Section~\ref{sec:exp}, Tables~\ref{tab:milan} and~\ref{tab:cross}).} We build a unified pipeline over three real datasets (Milan CDR, Shanghai Telecom, YJMob100K) with 14 methods, including five recent IEEE baselines (2023--2026), and a three-way leakage audit. DSWM ranks first among non-ablated configurations on all three datasets.
\end{itemize}

A further motivation shapes our evaluation design. Comparisons of learning-based network controllers are often undermined by three silent leaks: random train/test splits that let the model memorize periodicity, normalization statistics computed on the full dataset, and evaluation protocols that differ across methods. We therefore fix temporal splits, derive all scaling factors from the training segment, give every method the identical observation set, and subject the full pipeline to a read-only leakage audit (Section~\ref{sec:setup}). We regard this protocol as part of the contribution, because the margin we report (0.109 over the strongest baseline on Milan) is meaningful only if the comparison is clean.

The remainder of the paper is organized as follows. Section~\ref{sec:related} reviews related work. Section~\ref{sec:system} presents the system model and problem formulation. Section~\ref{sec:method} details DSWM. Section~\ref{sec:exp} reports experiments. Section~\ref{sec:conclusion} concludes.

\section{Related Work}
\label{sec:related}

\subsection{UAV-BS Deployment and Trajectory Design}
Early work characterized the air-to-ground channel and derived optimal altitudes for coverage \cite{alhourani2014optimal,mozaffari2019tutorial}. Zeng and Zhang introduced the rotary-wing power model and optimized trajectories for energy efficiency \cite{zeng2017energy}. Optimization-based formulations then scaled to multi-UAV settings: Sun \textit{et al.} \cite{sun2024multi} minimized a weighted sum of delay, energy, and load imbalance via BCD with Karush--Kuhn--Tucker (KKT) bisection and successive convex approximation (SCA); Jia \textit{et al.} \cite{jia2025dynamic} combined K-means and Voronoi pre-deployment with whale optimization for hierarchical swarms. Related formulations addressed distributionally robust aerial edge computing \cite{jia2025distributionally}, wireless-powered data acquisition \cite{ning2024joint}, two-timescale mobile edge computing (MEC) scheduling \cite{sun2024tjcct}, sensing-communication-computing trajectories \cite{peng2024trajectory}, covert integrated sensing and communication (ISAC) beamforming \cite{deng2024joint,cheng2025networked}, and self-adjusting slicing \cite{li2025self}; learning-based variants replace the solver with a policy network \cite{bai2025dynamic,hoang2025adaptive,liu2024deep,liu2025energy,chen2025spatiotemporal,meer2025hierarchical,peng2020multi,chen2022deep}. The gap in this line is per-slot re-solving against a demand snapshot: when demand drifts between solves, BCD-type methods are the most sensitive in our benchmark (Section~\ref{sec:ood}).

\subsection{Spatio-Temporal Demand Prediction}
Cellular and mobility demand prediction spans graph networks, Transformers, and large language models: MVSTGN fused multi-view spatial-temporal graphs for cellular traffic \cite{yao2021mvstgn}; Gu \textit{et al.} predicted city-level traffic with a spatial-temporal Transformer \cite{gu2023spatial}; probabilistic graph networks quantified travel demand uncertainty \cite{wang2024uncertainty}; recent work used confounder representations \cite{ji2025seeing}, traffic LLMs \cite{liu2025stllm}, and diffusion-based text-to-traffic generation \cite{zhang2024chattraffic}. These models output a forecast that a downstream planner must trust; the gap is objective misalignment, since per-cell forecasting accuracy does not measure demand-weighted coverage. We keep prediction inside a decision loop and show that decision-time use of observations, not forecast skill, is the binding constraint.

\subsection{World Models and Decision-Time Planning}
World models learn latent dynamics and plan by imagination \cite{ha2018world,hafner2019learning}; TD-MPC combined model predictive control with temporal-difference learning \cite{hansen2022temporal}; the Dreamer line scaled world-model agents across domains \cite{hafner2020mastering,hafner2025mastering,hafner2025training}. Self-supervised objectives \cite{balestriero2025lejepa,assran2025vjepa2,zhou2024dinowm} stabilize representation learning without reconstruction heuristics, and general agents have been argued to require world models \cite{richens2025general}. In wireless, MobiWorld built world models of mobile networks for traffic and channel simulation \cite{chai2025mobiworld}, and internal-inference frameworks moved learning inside the network \cite{ding2024from}. The gap is that existing wireless world models target prediction or simulation, not closed-loop repositioning with a physical service model inside the imagination. Offline model-based RL work such as MOPO \cite{yu2020mopo} motivates our uncertainty penalty, which we apply to imagined rollouts rather than policy training.

\subsection{Closest Works and Differentiation}
Four works are closest to ours, and the distinctions are structural. GA-MATR \cite{feng2024graph} learns a graph-attention MARL policy with a fairness regularizer; it is model-free, so it cannot imagine counterfactual demand, yet it is the strongest learning baseline under distribution shift in our benchmark (Section~\ref{sec:ood}), though still far below DSWM on weekdays. HRL-TPRA \cite{yuan2025hierarchical} decomposes the problem hierarchically in time with pointer-network trajectory planning and per-UAV resource actors; the hierarchy is compute-bound and ranked last among learning baselines in our setting (0.485). JTORATC \cite{sun2024multi} solves a weighted BCD problem per decision window; it carries no learned demand model and suffers OOD backfire (holiday 0.495 vs.\ weekday 0.583). MobiWorld \cite{chai2025mobiworld} builds a generative world model of wireless networks but does not close the loop with a service-physics-aware planner. DSWM differs from all four in that the service physics is differentiable inside the world model, so gradients and planning scores both flow through the true coverage mechanism.

\section{System Model and Problem Formulation}
\label{sec:system}
\subsection{Scenario and Assumptions}
We consider a city-wide aerial access layer. A fleet of $K=16$ UAV-BSs hovers at altitude $h_{\mathrm{fly}}=100$~m over a service area rasterized into $G$ cells ($G{=}400$ for Milan) of side $c$ (235~m for Milan). Time is slotted into $T_{\mathrm{slot}}=10$~min slots; one episode is one day of $T=144$ slots. The demand in slot $t$ is a nonnegative map $x_t \in \mathbb{R}_+^G$, normalized by the 99th percentile of the training segment. We adopt the following assumptions, each of which matches the deployed environment used in all experiments.

\begin{itemize}
    \item[(A1)] \textbf{Fixed flight altitude.} All UAV-BSs operate at a constant altitude of $h_{\mathrm{fly}}=100$~m. This isolates the horizontal repositioning problem, as optimal altitude results are well-established \cite{alhourani2014optimal}.
    
    \item[(A2)] \textbf{UAV mobility.} A UAV travels at most $D_{\max}=1500$~m per 10-minute slot. This fits comfortably within the $V_{\max}=25$~m/s speed limit of consumer-grade platforms, requiring an average speed of only 2.5~m/s.
    
    \item[(A3)] \textbf{Nearest-available association.} Each cell is served by the closest active UAV-BS. This practical Voronoi-style coverage \cite{jia2025dynamic} simplifies the model by decoupling user association from the learning problem.
    
    \item[(A4)] \textbf{Homogeneous rate demand.} Each normalized demand unit corresponds to $r_0=0.5$~Mbps per user. This isolates coverage optimization from traffic heterogeneity, which is typically handled by separate prediction modules \cite{yao2021mvstgn}.
    
    \item[(A5)] \textbf{Battery constraints.} Since finite endurance dictates short-horizon planning \cite{zeng2017energy}, we explicitly model charging cycles. A UAV dropping below a 15\% reserve returns to a quadrant station, charges for 6 slots (1 hour), and redeploys.
\end{itemize}

Table~\ref{tab:notation} lists the main notation.

\begin{table}[!t]
\caption{Main Notation}
\label{tab:notation}
\centering
\scriptsize
\setlength{\tabcolsep}{2pt}
\label{tab:notation}
\centering
\begin{tabular}{ll}
\toprule
Symbol & Meaning \\
\midrule
$t$, $T$ & slot index; slots per day ($T{=}144$) \\
$T_{\mathrm{slot}}$ & slot duration (10 min) \\
$k$, $K$ & UAV-BS index; fleet size ($K{=}16$) \\
$g$, $G$ & cell index; number of cells ($G{=}400$) \\
$c$ & cell side length (235/470/500~m) \\
$x_t \in \mathbb{R}_+^G$ & normalized demand map \\
$s_t^k=(p_t^k,b_t^k,\xi_t^k)$ & UAV $k$ state (position, battery, charge) \\
$a_t \in [-1,1]^{K \times 3}$ & action $(\Delta x,\Delta y,\mathrm{bw\_logit})$ \\
$h_{\mathrm{fly}}$, $D_{\max}$, $V_{\max}$ & altitude; move limit; speed limit \\
$B$, $N_0$, $f_c$ & bandwidth; noise spectral density; carrier frequency \\
$\gamma$, $E$ & planner discount; ensemble size \\
$N_{\mathrm{cem}}$, $N_{\mathrm{elite}}$ & CEM samples; elites \\
$m$, $\sigma_0$ & CEM sampling mean; initial sampling std \\
$\nu$ & EMA coefficient of the teacher \\
$o_t$ & observation (past $w{=}6$ demand, fleet, time) \\
$h_t$, $z_t$ & GRU deterministic / stochastic latent state \\
$q_\phi$, $p_\theta$ & posterior / prior of the RSSM \\
$\SR_t \in [0,1]$ & served ratio (aggregate demand satisfaction) \\
$r_0$, $U_{\mathrm{peak}}$ & per-user rate demand; satisfaction threshold \\
$E_t$, $E_{\mathrm{hover}}$, $E_{\mathrm{batt}}$ & slot / hover / battery energy \\
$P_{\mathrm{los}}$ & LoS probability \\
$R_{g,k}$ & Shannon rate of link $(g,k)$ \\
$H$, $\rho$ & horizon; anchor mixing coefficient \\
$u_t$, $\lambda_u$ & ensemble uncertainty; penalty weight \\
$r_s$ & Spearman rank correlation \\
\bottomrule
\end{tabular}
\end{table}

\subsection{Channel Model}
Let $d_{g,k}=\sqrt{h_{\mathrm{fly}}^2 + \lVert p_t^k - c_g \rVert^2}$ be the distance between UAV $k$ at position $p_t^k$ in slot $t$ and the center $c_g$ of cell $g$, and let $\theta_{g,k}=\arcsin(h_{\mathrm{fly}}/d_{g,k})$ be the elevation angle, converted to degrees before evaluating \eqref{eq:plos}. The LoS probability follows the standard sigmoid model \cite{alhourani2014optimal,mozaffari2019tutorial}:
\begin{equation}
P_{\mathrm{los}}(\theta_{g,k}) = \frac{1}{1 + a_{\mathrm{env}} \exp\!\big(-b_{\mathrm{env}}(\theta_{g,k} - a_{\mathrm{env}})\big)},
\label{eq:plos}
\end{equation}
where $a_{\mathrm{env}}$ and $b_{\mathrm{env}}$ are environment constants. The expected path loss in dB mixes LoS and non-LoS (NLoS) components with excess losses $\eta_{\mathrm{LoS}}$ and $\eta_{\mathrm{NLoS}}$; the constants $a_{\mathrm{env}}$, $b_{\mathrm{env}}$ and the excess-loss terms $\eta_{\mathrm{LoS}}$, $\eta_{\mathrm{NLoS}}$ follow the urban parametrization of \cite{alhourani2014optimal}:
\begin{align}
\mathrm{PL}_{g,k} = {}& 20\log_{10}\!\Big(\frac{4\pi f_c d_{g,k}}{c_0}\Big) \nonumber\\
& + P_{\mathrm{los}}(\theta_{g,k})\,\eta_{\mathrm{LoS}} + \big(1-P_{\mathrm{los}}(\theta_{g,k})\big)\,\eta_{\mathrm{NLoS}},
\label{eq:pl}
\end{align}
with carrier frequency $f_c=2$~GHz and light speed $c_0$. With transmit power $P_k=0.5$~W, allocated bandwidth $B_{g,k}$, and noise spectral density $N_0=10^{-20.4}$~W/Hz, the Shannon rate of link $(g,k)$ is
\begin{equation}
R_{g,k} = B_{g,k}\log_2\!\Big(1 + \frac{P_k\,10^{-\mathrm{PL}_{g,k}/10}}{N_0 B_{g,k}}\Big),
\label{eq:shannon}
\end{equation}
where $\sum_g B_{g,k} \le B$ with $B=20$~MHz per UAV-BS. The chain \eqref{eq:plos}--\eqref{eq:shannon} is deterministic given positions and is differentiable almost everywhere; Section~\ref{sec:decomp} reuses it inside the world model.

\subsection{Energy Consumption and Battery State Machine}
Rotor power follows the model of Zeng and Zhang \cite{zeng2017energy}:
\begin{align}
P(V) = {}& P_0\Big(1 + \frac{3V^2}{U_{\mathrm{tip}}^2}\Big)
+ P_i\Big(\sqrt{1 + \frac{V^4}{4v_0^4}} - \frac{V^2}{2v_0^2}\Big)^{1/2} \nonumber\\
& + \tfrac{1}{2} d_0 \varrho_{\mathrm{air}} s A V^3,
\label{eq:zeng}
\end{align}
where $V$ is speed, $P_0$ and $P_i$ are blade profile and induced power in hover, $U_{\mathrm{tip}}$ is tip speed, $v_0$ is mean rotor induced velocity, and $d_0$, $\varrho_{\mathrm{air}}$, $s$, $A$ are fuselage drag ratio, air density, rotor solidity, and disc area ($\rho$ is reserved for the anchor mixing coefficient). The per-UAV slot energy is $E_t^k = P(V_t^k)\,T_{\mathrm{slot}}$, and $E_t = \sum_k E_t^k$ is the fleet total; hover energy $E_{\mathrm{hover}} = P(0)\,T_{\mathrm{slot}}$ is the reference. Because repositioning spans at most 1500~m per 10-min slot, hovering dominates consumption; this physical fact explains the narrow energy spread ($\sim$3\%) across methods in Section~\ref{sec:main}. The battery obeys $b_{t+1}^k = b_t^k - E_t^k$ while airborne; when $b_t^k < 0.15\,E_{\mathrm{batt}}$ with $E_{\mathrm{batt}} = 1{,}972{,}800$~J (548~Wh, DJI M300-class), UAV $k$ enters \textit{returning}, flies to its quadrant station, charges for 6 slots, and redeploys at full charge.

\subsection{Demand-Service Model}
Cell $g$ requests an aggregate rate $x_t^g r_0$. Under nearest-available association (A3), the served demand of cell $g$ in slot $t$ is
\begin{equation}
y_t^g = \min\!\big(x_t^g r_0,\; R_{g,k^*(g)}\big),\qquad
k^*(g) = \arg\min_{k \in \mathcal{A}_t} \lVert p_t^k - c_g \rVert,
\label{eq:serve}
\end{equation}
where $\mathcal{A}_t$ is the set of airborne, non-returning UAV-BSs and $R_{g,k}$ is from \eqref{eq:shannon}. The served ratio aggregates fulfilled demand over the map:
\begin{equation}
\SR_t = \frac{\sum_g y_t^g}{\sum_g x_t^g r_0}
= \frac{\sum_g x_t^g \min\!\big(1,\; R_{g,k^*(g)}/(x_t^g r_0)\big)}{\sum_g x_t^g}.
\label{eq:sr}
\end{equation}
$\SR_t$ is demand-weighted, so covering a hotspot matters more than covering a sparse cell. A cell counts as satisfied when its fulfillment ratio reaches $U_{\mathrm{peak}}=0.8$; this threshold is used only to mark satisfied cells in the Jain fairness statistic reported in Section~\ref{sec:main}. The reward used by learning methods is
\begin{equation}
\mathcal{R}_t = \SR_t - \beta\,(1-\SR_t) - \alpha\,\frac{E_t}{E_{\mathrm{hover}} K},
\label{eq:reward}
\end{equation}
with $\alpha=\beta=1.0$ in all experiments.

\subsection{Problem Formulation}
The operator seeks a policy $\pi$ mapping observations to actions that maximizes cumulative served ratio over an episode:
\begin{align}
\mathrm{(P1):}\quad \max_{\pi}\; & \mathbb{E}\Big[\textstyle\sum_{t=1}^{T} \SR_t\Big] \label{eq:p1}\\
\mathrm{s.t.}\; & \lVert \Delta p_k^t \rVert \le D_{\max},\; \lVert \Delta p_k^t \rVert/T_{\mathrm{slot}} \le V_{\max}, \tag{C1}\label{eq:c1}\\
& b_{t+1}^k = b_t^k - E_t^k,\; b_t^k < 0.15\,E_{\mathrm{batt}} \Rightarrow \text{return}, \tag{C2}\label{eq:c2}\\
& \textstyle\sum_{g} B_{g,k} \le B,\; \forall k,t. \tag{C3}\label{eq:c3}
\end{align}
Constraint \eqref{eq:c1} enforces kinematics, \eqref{eq:c2} the battery state machine, and \eqref{eq:c3} the per-UAV bandwidth budget. The observation $o_t$ contains the past $w=6$ demand maps, the current demand map, fleet position/battery/charge states, and a time-phase encoding; all methods in Section~\ref{sec:exp} share this identical information set.

(P1) is hard for three reasons, and each maps to one DSWM component. First, the demand map is time-varying and partially observed through a short window, so per-slot re-optimization chases a moving target; DSWM answers this with a learned world model that carries demand dynamics in latent state. Second, the objective couples a nonconvex air-to-ground channel with association and battery constraints, so the composite problem is nonconvex and gradients through solvers are unavailable; DSWM answers this with a decomposed differentiable service head that makes the coverage mechanism differentiable. Third, credit assignment spans slots: a repositioning decision pays off hours later, after charging and drift; DSWM answers this with CEM planning over a horizon of $H=15$ slots (2.5~h) anchored on the current observation.

\begin{figure*}[!t]
\centering
\includegraphics[width=0.9\textwidth]{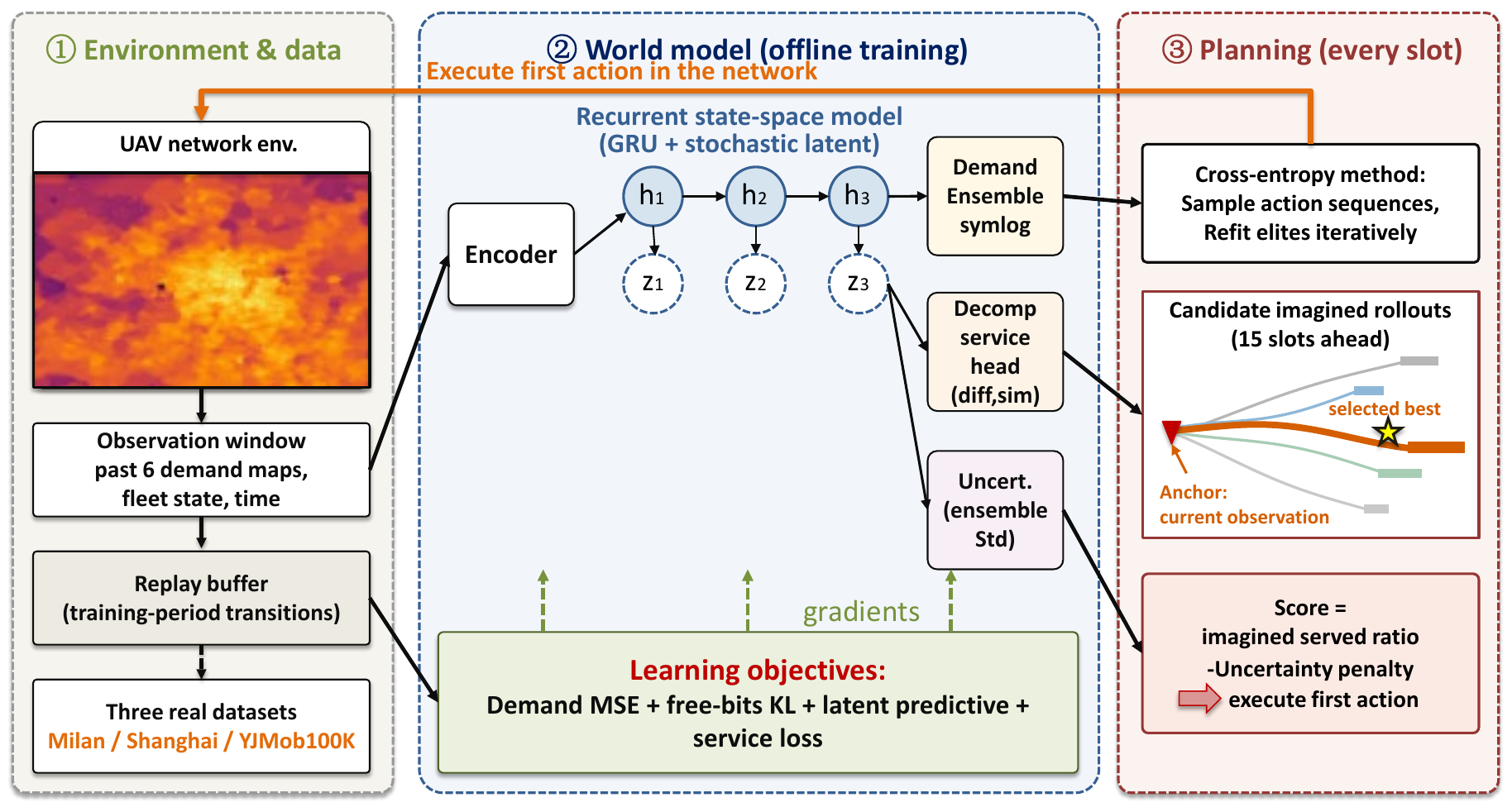}
\caption{DSWM framework. The encoder and RSSM maintain a latent state from the observation window. The CEM planner imagines $H$-slot rollouts, anchors the imagined demand on the current observation with mixing coefficient $\rho=0.95$, scores each rollout with the decomposed service head minus an uncertainty penalty, and executes the first action. Training (dashed) is offline over replayed sequences.}
\label{fig:framework}
\end{figure*}

\section{Proposed Method: DSWM}
\label{sec:method}

\subsection{Overview}
Fig.~\ref{fig:framework} shows the data flow. At each slot, the encoder compresses the observation $o_t$ and the RSSM updates its latent state $(h_t, z_t)$. The CEM planner samples candidate action sequences, rolls them forward for $H$ slots inside the world model, converts each imagined latent state into a served-ratio estimate through the decomposed service head, penalizes ensemble disagreement, and executes the first action of the best sequence. Training is offline: sequences from a replay buffer of pre-collected transitions update the RSSM, the demand ensemble, the latent predictive objective, and the service head end to end. The same information set ($o_t$ and internal states) is available to DSWM and to every baseline.

\subsection{RSSM World Model with Latent Predictive Regularization}
The world model is an RSSM \cite{hafner2019learning,hafner2025mastering} with a deterministic GRU (gated recurrent unit) state $h_t$ and a diagonal-Gaussian stochastic state $z_t$:
\begin{align}
h_t &= f_\theta(h_{t-1}, z_{t-1}, a_{t-1}), \label{eq:gru}\\
\text{prior:}\quad p_\theta(z_t \mid h_t) &= \mathcal{N}\big(\mu_\theta(h_t), \sigma_\theta^2(h_t)\big), \label{eq:prior}\\
\text{posterior:}\quad q_\phi(z_t \mid h_t, o_t) &= \mathcal{N}\big(\mu_\phi(h_t, e_\psi(o_t)), \label{eq:post}\\
 &\quad\ \sigma_\phi^2(h_t, e_\psi(o_t))\big), \nonumber
\end{align}
where $e_\psi$ is the observation encoder. Imagination uses the prior; training uses the posterior. The Kullback--Leibler (KL) term uses free bits with the Dreamer-style balancing weights $0.8/0.2$ \cite{hafner2025mastering}:
\begin{equation}
L_{\mathrm{KL}} = 0.8\,\KL\big(\sg[q_\phi] \,\|\, p_\theta\big) + 0.2\,\KL\big(q_\phi \,\|\, \sg[p_\theta]\big),
\label{eq:kl}
\end{equation}
where each per-dimension KL is clipped from below by a free-bits floor and $\sg[\cdot]$ stops gradients; the $0.8$-weighted term trains the prior toward the stopped posterior, and the $0.2$-weighted term trains the posterior toward the stopped prior. Free bits prevent posterior collapse, which we found to be the main stability risk during training scans.

Two implementation choices: the demand ensemble operates in symlog space, $\mathrm{symlog}(x)=\mathrm{sign}(x)\ln(1+|x|)$, because demand is heavy-tailed and a linear-scale loss would fit the mean and ignore hotspots; and a GRU deterministic path instead of a Transformer backbone, since the observation window is short ($w=6$ slots), the planning-relevant state is Markovian given $h_t$, and recurrence keeps the imagination forward pass cheap for CEM.
Reconstruction-free shaping follows the joint-embedding predictive (JEPA) view \cite{assran2025vjepa2}. Instead of reconstructing the demand map (which forces the model to spend capacity on heavy-tailed magnitudes irrelevant to coverage), DSWM predicts in representation space. Let $\hat h_{t+1} = g_\omega(h_t)$ be the one-step prediction of the recurrent latent; the target $\bar h_{t+1}$ is produced by an exponential-moving-average (EMA) teacher, a delayed copy of the encoder--RSSM stack whose parameters $\bar\theta$ track the online parameters $\theta$ after every gradient step by
\begin{equation}
\bar\theta \leftarrow \nu\,\bar\theta + (1-\nu)\,\theta,
\label{eq:emaupdate}
\end{equation}
and $\bar h_{t+1}$ denotes the recurrent latent produced by this teacher. The latent prediction loss is
\begin{equation}
L_{\mathrm{pred}} = \big\lVert g_\omega(h_t) - \sg[\bar h_{t+1}] \big\rVert_2^2,
\label{eq:ema}
\end{equation}
where $g_\omega$ is a latent predictor conditioned on the recurrent state and $\nu$ is the EMA coefficient. Thus \eqref{eq:ema} and the first term of \eqref{eq:lpr} are the same loss, written once and reused, not two different objectives. A joint-embedding loss alone admits the collapsing solution $f_\theta \equiv \mathrm{const}$, which makes \eqref{eq:ema} identically zero but destroys all information. We prevent collapse with a per-dimension variance margin in the style of VICReg \cite{chen2022intra},
\begin{equation}
\begin{split}
L_{\mathrm{LPR}} = \mathbb{E}\,\big\lVert \hat h_{t+1} - \sg[\bar h_{t+1}] \big\rVert_2^2 \\
+ \lambda_{\mathrm{var}} \sum_j \max\big(0,\, m_{\mathrm{var}} - \mathrm{Std}_j(h)\big),
\end{split}
\label{eq:lpr}
\end{equation}
where the first term instantiates \eqref{eq:ema} on the recurrent latent ($\hat h_{t+1}=g_\omega(h_t)$, $\bar h_{t+1}$ the EMA target), $\mathrm{Std}_j$ is taken over the batch dimension for latent dimension $j$, and $m_{\mathrm{var}}$ is the variance margin.

This objective replaced an earlier reconstruction term after an internal scan: it raised the weekday served ratio from 0.878 to 0.888 and stabilized the OOD split. The total world-model loss is
\begin{equation}
L_{\mathrm{WM}} = L_{\mathrm{dem}} + \beta_{\mathrm{KL}}\,L_{\mathrm{KL}} + \lambda_L\,L_{\mathrm{LPR}} + L_{\mathrm{serve}},
\label{eq:lwm}
\end{equation}
where $L_{\mathrm{dem}}$ is the demand-head loss and $L_{\mathrm{serve}}$ is the service-head loss defined next.
\subsection{Decomposed Demand-Service Physics Head}
\label{sec:decomp}
\begin{figure*}[!t]
\centering
\includegraphics[width=1.6\columnwidth]{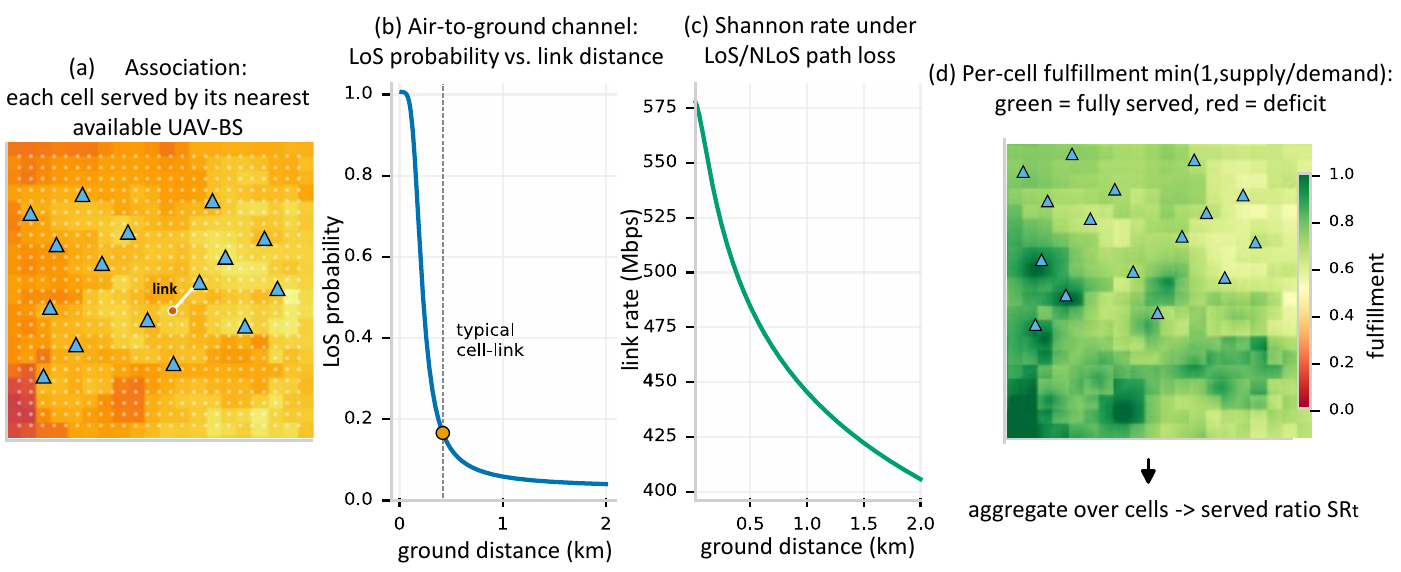}
\caption{Decomposed demand-service physics head. An $E$-head ensemble predicts the next demand map in symlog space; a differentiable service simulator replays association $\rightarrow$ probabilistic LoS \eqref{eq:plos} $\rightarrow$ Shannon rate \eqref{eq:shannon} $\rightarrow$ capped per-cell fulfillment $\rightarrow$ aggregate $\widehat{\SR}$ \eqref{eq:srhat}. Ensemble disagreement defines the uncertainty $u_t$.}
\label{fig:decomp}
\end{figure*}

Fig.~\ref{fig:decomp} illustrates the service-physics chain that the head replays differentiably---association, probabilistic LoS channel, Shannon rate, and capped per-cell fulfillment. A demand ensemble of $E$ heads predicts the next-slot demand map in symlog space, $\hat x_{t+1}^{(e)} = \mathrm{symlog}^{-1}\big(D_e(h_{t+1}, z_{t+1})\big)$, and their mean $\hat x_{t+1}$ is the demand estimate. The head then replays the service physics of Section~\ref{sec:system} as a differentiable simulator: given imagined UAV positions decoded from the action sequence, it computes association by nearest available UAV, evaluates \eqref{eq:plos}--\eqref{eq:shannon} for each link, caps per-cell fulfillment, and aggregates:
\begin{equation}
\widehat{\SR}_{t+1} = \frac{\sum_g \hat x_{t+1}^g \min\!\big(1,\, \hat R_{g,k^*(g)}/(\hat x_{t+1}^g r_0)\big)}{\sum_g \hat x_{t+1}^g}.
\label{eq:srhat}
\end{equation}
Because every step of \eqref{eq:srhat} is differentiable, gradients flow through the coverage mechanism back into the world model; the model is therefore trained on the quantity the operator cares about, not on a proxy. The service loss is $L_{\mathrm{serve}} = \mathbb{E}\,(\widehat{\SR}_{t+1} - \SR_{t+1})^2$ on replayed transitions. Removing this decomposition and regressing $\SR_t$ directly from latent state costs $0.120$ of served ratio (Section~\ref{sec:ablation}), which confirms that the physics, not the latent capacity, carries the coverage signal.

Ensemble disagreement defines the uncertainty of an imagined step:
\begin{equation}
u_{t+1} = \mathrm{Std}_e\big(\hat{x}_{t+1}^{(e)}\big) \;\text{aggregated over cells},
\label{eq:unc}
\end{equation}
and the planner subtracts $\lambda_u u_\tau$ from imagined rewards ($\lambda_u = 1.0$ in production), following the pessimism principle of offline model-based RL \cite{yu2020mopo}. Removing this term changes the score by only $+0.004$ (Section~\ref{sec:ablation}): the uncertainty penalty acts as insurance against rare overconfident branches, not as a gain source.

\subsection{Observation-Anchored CEM Decision-Time Planning}
\label{sec:planning}
\begin{figure}[!t]
\centering
\includegraphics[width=0.8\columnwidth]{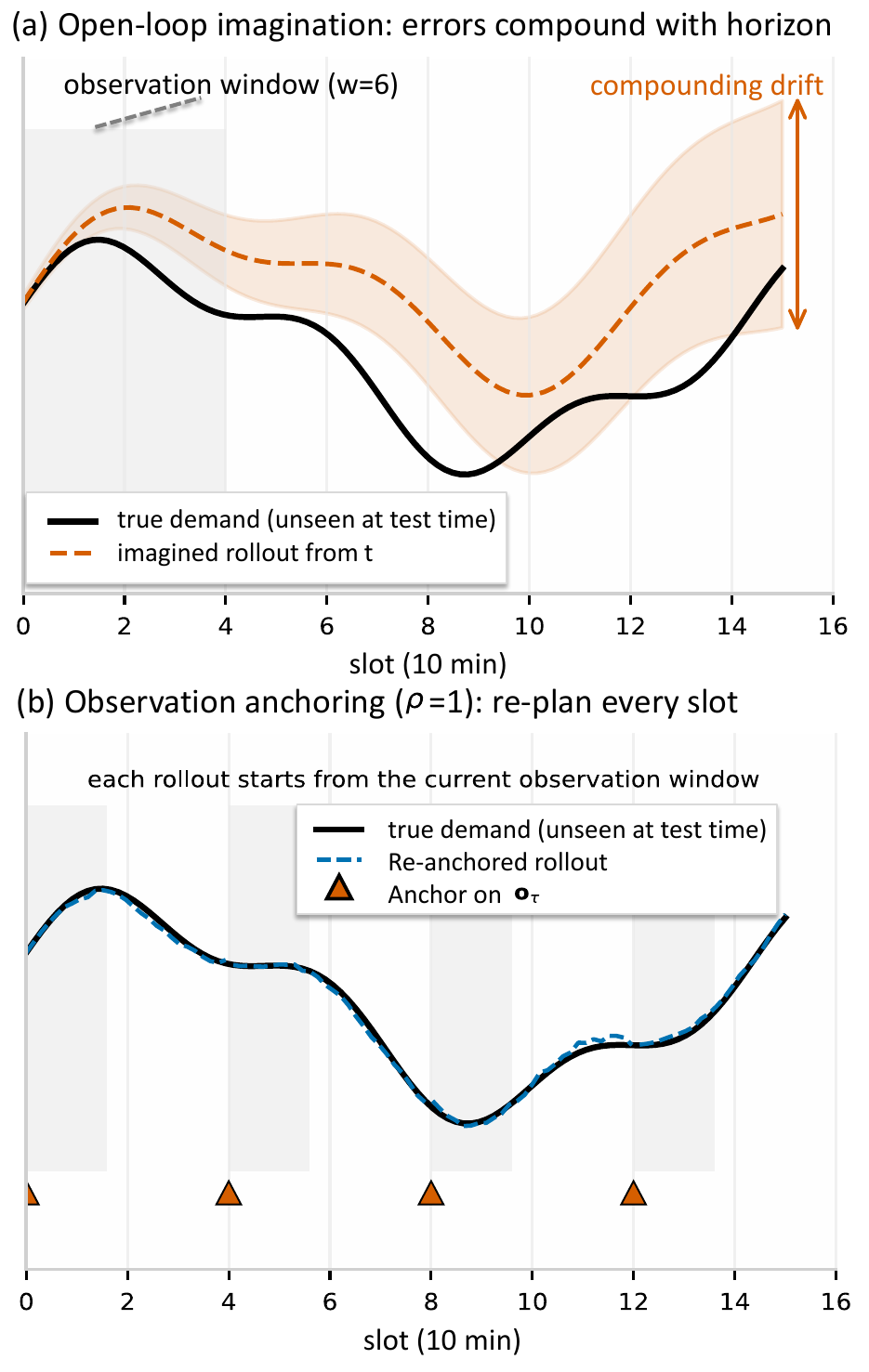}
\caption{Observation-anchored CEM planning. Candidate action sequences are imagined for $H=15$ slots through the RSSM prior; at each imagined step the demand estimate is anchored on the current observation with mixing coefficient $\rho=0.95$ \eqref{eq:anchor}; rollouts are scored by $J$ in \eqref{eq:cemscore}; elites refit the sampling distribution and the first action is executed.}
\label{fig:anchor}
\end{figure}

Fig.~\ref{fig:anchor} illustrates the mechanism behind the planner: open-loop imagination accumulates error along the horizon, while re-anchoring each rollout to the current observation keeps the estimate grounded. At slot $t$, the planner samples $N_{\mathrm{cem}}$ action sequences $a_{t:t+H-1}$ from a diagonal Gaussian, rolls each through the prior \eqref{eq:gru}--\eqref{eq:prior}, and scores the rollout as
\begin{equation}
J\big(a_{t:t+H-1}\big) = \sum_{\tau=t+1}^{t+H} \gamma^{\tau-t}\,\big(\widehat{\SR}_\tau - \lambda_u u_\tau\big),
\label{eq:cemscore}
\end{equation}
where $H=15$ slots (2.5~h) and $\gamma$ is the planner discount. The top $N_{\mathrm{elite}}$ sequences refit the sampling distribution, and after a fixed number of iterations the first action of the best sequence is executed.

The key mechanism is observation anchoring. With mixing coefficient $\rho$, the imagined demand at each rollout step is blended with the current observed demand map (persist anchoring):
\begin{equation}
\hat x_\tau \leftarrow \rho\, x_t + (1-\rho)\,\hat x_\tau .
\label{eq:anchor}
\end{equation}
Equation \eqref{eq:anchor} is a deterministic convex interpolation; we use $\rho=0.95$, so the estimate leans on the current observation with a small forecast weight. Even so, the score $J$ still reflects multi-slot consequences that no myopic rule can see---fleet positions, battery depletion, forced returns, and charging occupancy propagated through the service simulator---while compounding demand error along the horizon is removed. In a preliminary smaller-scale setting ($K{=}4$, without the battery model), this choice improved both splits by $+0.018$ over the non-anchored variant at no cost (replicated from two checkpoints), and demand-forecasting enhancements were net negative; the interpretation is that the current observation already carries most of the actionable signal, so the model evaluates consequences of motions under it rather than out-forecasting the environment. Algorithm~\ref{alg:cem} gives the per-slot procedure; an actor/critic direct mode exists as a fallback but is not used in the reported results.

\begin{algorithm}[!t]
\small
\caption{Per-Slot Observation-Anchored CEM Planning}
\label{alg:cem}
\begin{algorithmic}[1]
\REQUIRE observation $o_t$; latent state $(h_{t-1}, z_{t-1})$; horizon $H$; anchor $\rho=0.95$; samples $N_{\mathrm{cem}}$; elites $N_{\mathrm{elite}}$; iterations $I$
\STATE Encode $o_t$ and update posterior: $h_t, z_t \leftarrow$ \eqref{eq:gru},\eqref{eq:post}
\STATE Initialize sampling mean $m \leftarrow 0$, std $\Sigma \leftarrow \sigma_0^2\,\mathbf{I}$ over $a_{t:t+H-1}$
\FOR{$i = 1$ \TO $I$}
  \STATE Sample $N_{\mathrm{cem}}$ sequences $\{a_{t:t+H-1}^{(n)}\} \sim \mathcal{N}(m, \Sigma)$
  \FOR{each sequence $n$}
    \STATE Roll out $(h_\tau, z_\tau)$ for $H$ steps via \eqref{eq:gru},\eqref{eq:prior}
    \STATE Predict $\hat x_\tau^{(e)}$, anchor by \eqref{eq:anchor} with $\rho=0.95$
    \STATE Evaluate $\widehat{\SR}_\tau$ by \eqref{eq:srhat} and $u_\tau$ by \eqref{eq:unc}
    \STATE Score $J^{(n)}$ by \eqref{eq:cemscore}
  \ENDFOR
  \STATE Select $N_{\mathrm{elite}}$ top sequences; refit $m, \Sigma$ to elites
\ENDFOR
\STATE \textbf{return} first action $a_t^{*}$ of the best-scoring sequence
\end{algorithmic}
\end{algorithm}

\subsection{Training Procedure}
Algorithm~\ref{alg:train} summarizes offline training. A heuristic pre-collection phase gathers 60{,}000 transitions on training days only; the replay buffer never sees test-period data. Training then runs 30{,}000 gradient steps (learning rate $10^{-3}$), which an internal scan showed is well past the plateau (curves flatten near 20k steps). Each step samples sequences, encodes them, unrolls the RSSM, and applies the four loss terms of \eqref{eq:lwm}: demand MSE in symlog space, free-bits KL \eqref{eq:kl}, latent predictive regularization \eqref{eq:lpr}, and the service loss, plus an uncertainty calibration term for the ensemble. The EMA teacher updates after each gradient step. The production configuration is identical across all evaluated datasets, with no per-dataset tuning.

\begin{algorithm}[!t]
\small
\caption{DSWM Offline Training}
\label{alg:train}
\begin{algorithmic}[1]
\REQUIRE training-days environment; collect steps $=60{,}000$; gradient steps $=30{,}000$; lr $=10^{-3}$
\STATE Pre-collect 60{,}000 transitions into replay (training days only)
\FOR{step $=1$ \TO $30{,}000$}
  \STATE Sample a batch of sequences from replay
  \STATE Encode observations; unroll RSSM via \eqref{eq:gru}--\eqref{eq:post}
  \STATE Demand ensemble: symlog MSE loss $L_{\mathrm{dem}}$ ($E$ heads, mean)
  \STATE Free-bits KL loss $L_{\mathrm{KL}}$ by \eqref{eq:kl} (weights 0.8/0.2)
  \STATE Latent predictive regularization $L_{\mathrm{LPR}}$ by \eqref{eq:lpr} (EMA teacher $+$ variance margin)
  \STATE Decomposed service loss $L_{\mathrm{serve}}$ via \eqref{eq:srhat}; uncertainty calibration via \eqref{eq:unc}
  \STATE Update parameters with $\nabla L_{\mathrm{WM}}$ \eqref{eq:lwm}; EMA-update teacher
\ENDFOR
\end{algorithmic}
\end{algorithm}

\subsection{Complexity Analysis}
At decision time, DSWM performs $N_{\mathrm{cem}} \times I$ imagined rollouts of $H$ steps each ($N_{\mathrm{cem}}$ candidates, $I$ elite-refit iterations, initial std $\sigma_0$; production values are hard-coded in the released \texttt{experiment.py} configuration). Every imagined step costs one GRU update, $E$ demand-head evaluations, and one pass of the differentiable service simulator, all of which are dense tensor operations that batch across the $N_{\mathrm{cem}}$ samples. Per-slot planning cost is therefore $O(N_{\mathrm{cem}} I H)$ imagination forwards with a small constant, and it does not grow with the nonconvexity of the problem. In contrast, the BCD baseline JTORATC \cite{sun2024multi} invokes an SCA solver per decision window; measured solve times exceeded 8~s per slot in our $K=16$ setting, which forced the documented degradation rule (one solve per 6 slots with reduced SCA iterations). DSWM's planning cost is fixed per slot and independent of solver convergence, which suits online repositioning.

\begin{figure*}[!t]
\centering
\includegraphics[width=1.6\columnwidth]{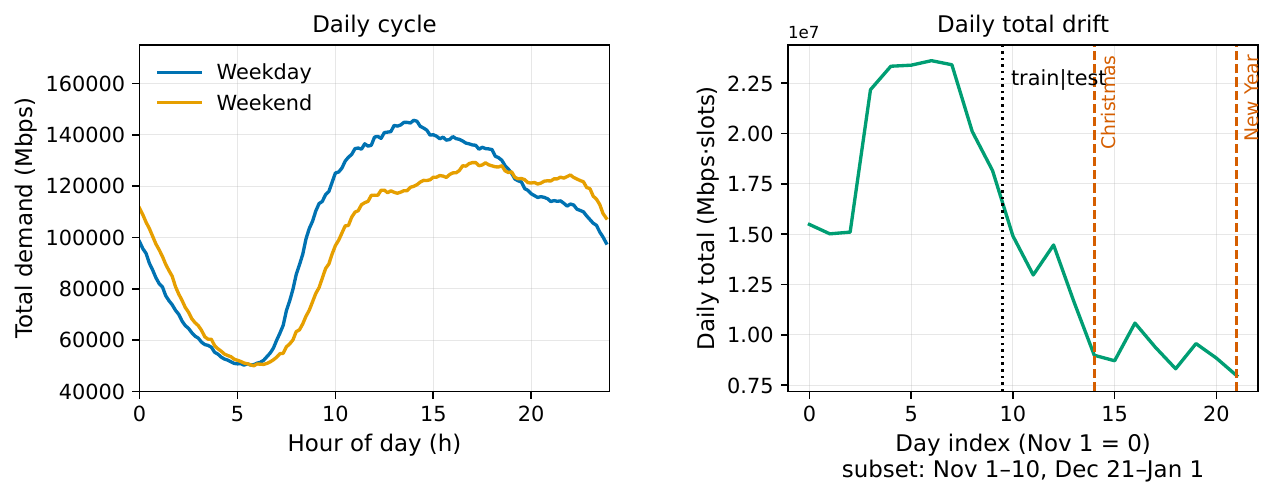}
\caption{Exploratory data analysis. Left: daily demand cycle (weekday vs.\ weekend; Milan shows a commute bimodal pattern). Right: daily total drift across the observation period, with the temporal train/test split and the Christmas--New Year OOD week marked. The spatial demand map is shown in Fig.~\ref{fig:scenario}. Milan demand is the most dispersed (top 5\% cells carry 13\% of demand, vs.\ 22\% for YJMob100K), and we show in Section~\ref{sec:cross} that this dispersion is what makes dynamic repositioning most valuable here.}
\label{fig:eda}
\end{figure*}

\begin{figure}[!t]
\centering
\includegraphics[width=0.8\columnwidth]{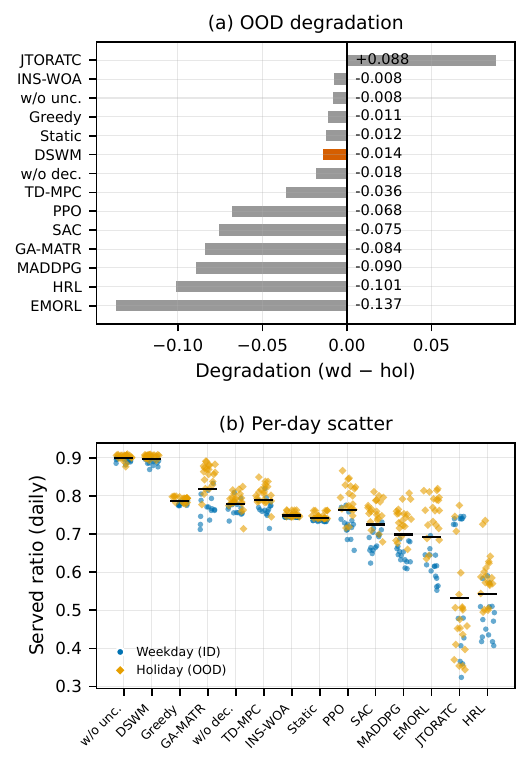}
\caption{OOD analysis over the test period. (a) per-method OOD degradation (weekday minus holiday); (b) per-day served-ratio scatter, including the holiday OOD week. DSWM is the most stable method and scores higher on holidays (0.903) than weekdays (0.889). GA-MATR attains the highest holiday score among learning baselines (0.854) and, together with TD-MPC, is one of only two learning baselines whose holiday score surpasses Greedy; JTORATC suffers OOD backfire (0.495 vs.\ 0.583). Degradation values are computed from unrounded per-day series.}
\label{fig:ood}
\end{figure}

\section{Experiments}
\label{sec:exp}

\subsection{Datasets and Preprocessing}
\label{sec:data}
We evaluate on three real datasets with one unified pipeline; Fig.~\ref{fig:eda} visualizes the demand statistics, and the splits are declared per dataset below.

\textit{Milan (main dataset).} The Telecom Italia CDR dataset \cite{barlacchi2015multi} (2013, internet modality), cropped to $20 \times 20$ cells of $c=235$~m. Training: Nov.\ 1--10 (10 days); testing: Dec.\ 21--Jan.\ 1 (12 days), containing the Christmas--New Year OOD week. Demand is normalized by the training-segment 99th percentile. Milan demand is extremely dispersed: the top 5\% cells carry only 13\% of demand, with a clear commute bimodal pattern.

\textit{Shanghai Telecom.} A public cellular dataset (2014-06--2014-11; 6.95M raw records, cleaned to 6.15M). Demand is the number of Internet session starts per cell per slot. Sparsity-adaptive preprocessing coarsens cells to $c=470$~m with a 3-slot sliding-window density estimate (zero-value rate 85.3\%, i.e., genuinely sparse sessions). 
Training: June--October (153 days, including the merged validation segment); testing: November (30 days). Weekends form the holiday (OOD) split.

\textit{YJMob100K.} The YJMob100K dataset \cite{yabe2024yjmob} (Zenodo 10142719, 2023), 111.5M rows. We declare the semantics explicitly: demand is the number of distinct user IDs present per cell per slot---an approximately 5\% sampled mobility proxy, not traffic bytes. Native 30-min slots are conservatively de-aggregated into 10-min slots (uniform within-cell division by 3, no information injected), and the densest $9 \times 9$ cells at $c=500$~m are cropped ($4.5 \times 4.5$~km, Milan-matched scale). Training: days 0--55; testing: days 56--74 (19 days). Week structure is recovered from autocorrelation (peak lag 7); pseudo-weekend labels (21 days) are inferred and declared as such. Intensity is aligned across datasets by scaling factors ($\kappa=3.81$ for Shanghai, $\kappa=1.749$ for YJMob100K).

\begin{figure*}[!t]
\centering
\includegraphics[width=1.98\columnwidth]{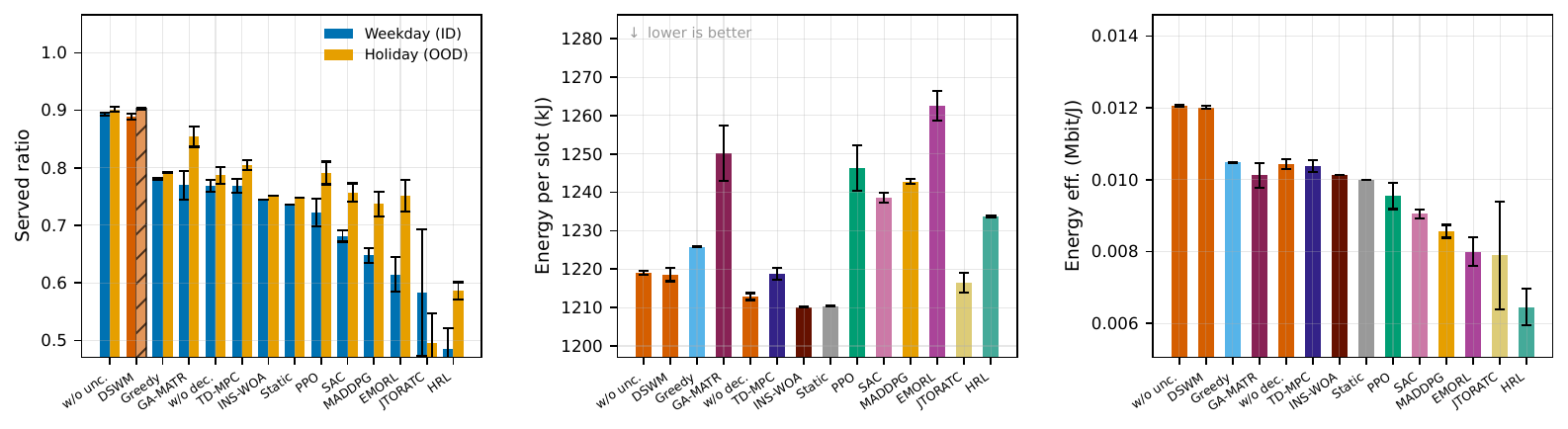}
\caption{Milan main results for all 14 methods: (a) served ratio (weekday ID / holiday OOD), (b) energy per slot, (c) energy efficiency. Served ratio is the only strongly discriminative metric; energy differs by at most $\sim$3\%. Fairness is reported in the text.}
\label{fig:main}
\end{figure*}

\subsection{Setup and Evaluation Protocol}
\label{sec:setup}
Table~\ref{tab:params} lists the scenario parameters with their justification. All learning methods receive the identical observation set (past $w=6$ demand maps, current demand, fleet position/battery/charge, time phase), and all actions are $(\Delta x, \Delta y, \mathrm{bw\_logit}) \in [-1,1]^{K \times 3}$.

The evaluation protocol is designed for credibility: temporal train/test splits (random splits forbidden); fixed evaluation days, weekday (in-distribution, ID) $\{0,1,2\}$ and holiday (OOD) $\{3,4,10,11\}$ (for Milan, Dec.\ 21--23 and Dec.\ 24--25/31/Jan.\ 1); 5 seeds on Milan and 3 per dataset. We report $\pm$std where the spread changes adjacent-ranking interpretation (DSWM, GA-MATR, JTORATC); per-day variability appears in Fig.~\ref{fig:ood}(b). 

Two engineering details support reproducibility: environment outputs drift by order $10^{-4}$ across days (BLAS/CPU variation), so all regression comparisons use frozen same-day references; and checkpoints are written only by the Milan full run (cross-dataset runs persist metrics only), so all reported numbers come from persisted per-seed result files, never from logs or probe curves.

\begin{table*}[!t]
\centering
\scriptsize

\begin{minipage}[b]{0.60\textwidth}
\captionof{table}{Scenario Parameters and Justification}
\label{tab:params}
\end{minipage}\hfill
\begin{minipage}[b]{0.385\textwidth}
\captionof{table}{Milan Main Results: Served Ratio (Weekday ID / Holiday OOD), 5 Seeds}
\label{tab:milan}
\end{minipage}

\vspace{4pt}

\begin{minipage}[t]{0.60\textwidth}
\centering
\setlength{\tabcolsep}{2pt}
\begin{tabular}[t]{@{}lll@{}}
\toprule
Parameter & Value & Basis \\
\midrule
Fleet size $K$ & 16 & city-wide; $\{2,8\}$ scaling \\
Altitude $h_{\mathrm{fly}}$ / speed $V_{\max}$ & 100 m / 25 m/s & low-altitude \cite{alhourani2014optimal} \\
Move limit $D_{\max}$ / slot & 1500 m / 10 min & per-slot motion limit \\
Carrier $f_c$ / bandwidth $B$ & 2 GHz / 20 MHz & sub-6GHz access \\
Power / noise $N_0$ & 0.5 W / $10^{-20.4}$ W/Hz & small-cell \\
Rate demand $r_0$ / threshold $U_{\mathrm{peak}}$ & 0.5 Mbps / 0.8 & per-user \\
Battery $E_{\mathrm{batt}}$ & 548 Wh & DJI M300-class \\
Reserve / recharge & 0.15 / 6 slots & forced return; 1 h charge \\
Charging stations & 4 & one per quadrant \\
Reward weights $\alpha,\beta$ & 1.0, 1.0 & equal coverage/energy \\
Obs.\ window $w$ / horizon $H$ & 6 / 15 slots & 1 h hist.\ / 2.5 h ahead \\
Anchor $\rho$ / uncertainty $\lambda_u$ & 0.95 / 1.0 & dual-ckpt; abl.\ Section~\ref{sec:ablation} \\
Learning rate & $10^{-3}$ & fixed across datasets \\
Gradient steps & 30{,}000 & 20k-step plateau observed \\
\bottomrule
\end{tabular}
\end{minipage}\hfill
\begin{minipage}[t]{0.385\textwidth}
\centering
\setlength{\tabcolsep}{1.5pt}
\begin{tabular}[t]{@{}llcc@{}}
\toprule
Method & Class & Weekday & Holiday \\
\midrule
\textbf{DSWM (ours)} & world model & \textbf{0.889$\pm$0.005} & \textbf{0.903} \\
w/o unc & ablation & 0.894 & 0.902 \\
w/o dec & ablation & 0.769 & 0.787 \\
Greedy (BCD) & model-free & 0.780 & 0.792 \\
GA-MATR \cite{feng2024graph} & MARL & 0.770$\pm$0.025 & 0.854 \\
TD-MPC \cite{hansen2022temporal} & model-based & 0.769 & 0.805 \\
INS-WOA \cite{jia2025dynamic} & optimization & 0.744 & 0.752 \\
Static (K-means) & deployment & 0.736 & 0.748 \\
PPO & model-free RL & 0.723 & 0.791 \\
SAC & model-free RL & 0.682 & 0.757 \\
MADDPG & MARL & 0.648 & 0.737 \\
EMORL-TCTO \cite{song2022evolutionary} & multi-obj.\ RL & 0.614 & 0.751 \\
JTORATC \cite{sun2024multi} & BCD optim. & 0.583$\pm$0.110 & 0.495 \\
HRL-TPRA \cite{yuan2025hierarchical} & hier.\ RL & 0.485 & 0.586 \\
\bottomrule
\end{tabular}
\end{minipage}
\end{table*}

\subsection{Baseline Lineup}
\label{sec:baselines}
Fourteen methods are compared. Seven are learning-based: PPO, SAC, MADDPG, TD-MPC \cite{hansen2022temporal}, and three state-of-the-art baselines re-implemented from recent works---GA-MATR \cite{feng2024graph} (graph-attention MARL with fairness regularizer), HRL-TPRA \cite{yuan2025hierarchical} (hierarchical trajectory planning with per-UAV resource actors), and EMORL-TCTO \cite{song2022evolutionary} (evolutionary multi-objective RL). Four are optimization-based or classical: JTORATC \cite{sun2024multi} (weighted BCD with KKT bisection and SCA, re-implemented), INS-WOA \cite{jia2025dynamic} (K-means/Voronoi pre-deployment with whale optimization, re-implemented), Static (K-means placement), and Greedy (per-slot BCD). Together with GA-MATR, HRL-TPRA, and EMORL-TCTO above, five IEEE baselines (2023--2026) are reproduced from their papers. DSWM and two internal ablations (w/o unc, w/o dec) complete the lineup. 

\subsection{Main Results}
\label{sec:main}

\begin{figure*}[!t]
\centering
\includegraphics[width=1.6\columnwidth]{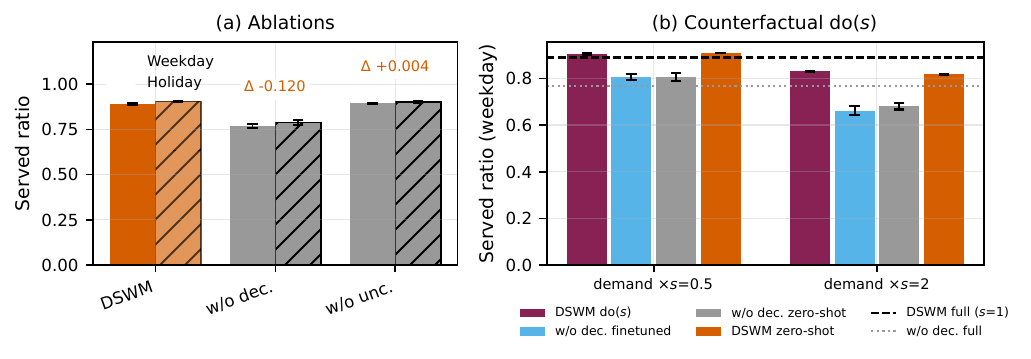}
\caption{Ablation and counterfactual analysis on Milan. Left: removing the decomposed service head costs $0.120$ of served ratio (0.769 vs.\ 0.889), while removing the uncertainty penalty gains $0.004$ (insurance, not a gain source). Right: counterfactual demand-scaling $\mathrm{do}(s)$ intervention (imagined demand scaled by $s$; dashed/dotted references at $s{=}1$).}
\label{fig:ablation}
\end{figure*}

\begin{figure*}[!t]
\centering
\includegraphics[width=1.6\columnwidth]{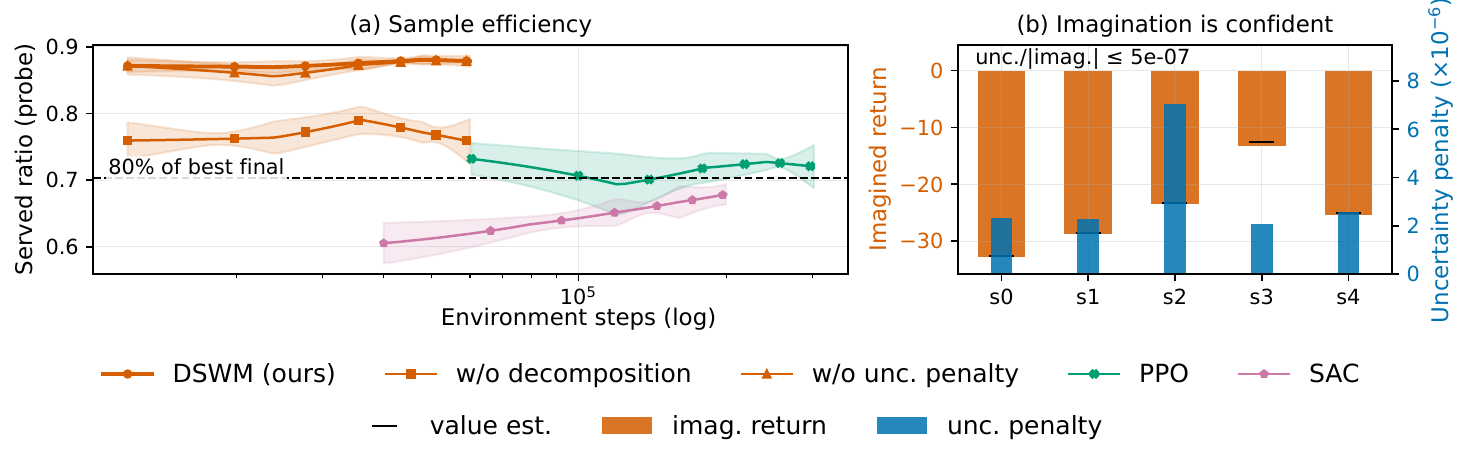}
\caption{Sample efficiency on Milan. Curves are probe-based and diagnostic only (fixed test-day probe; no early stopping or model selection). DSWM's probe flattens near 20k gradient steps; the production budget is 30k steps.}
\label{fig:sample}
\end{figure*}

\begin{figure*}[!t]
\centering
\includegraphics[width=1.6\columnwidth]{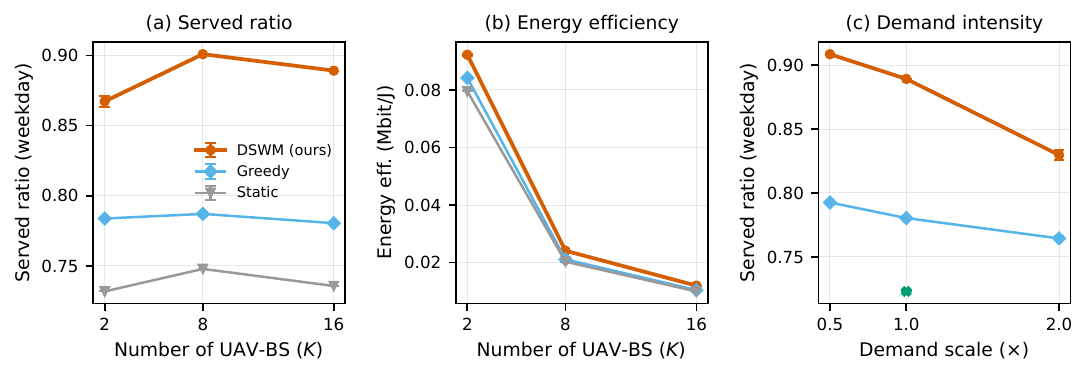}
\caption{Scalability in fleet size $K \in \{2,8,16\}$ and demand intensity on Milan. DSWM's advantage grows with fleet size, where coordination is hardest for myopic baselines.}
\label{fig:scaling}
\end{figure*}

Table~\ref{tab:milan} and Fig.~\ref{fig:main} report the Milan results. Two findings stand out.

First, DSWM ranks first among non-ablated configurations (the w/o-uncertainty variant is analyzed in Section~\ref{sec:ablation}) with $0.889\pm0.005$ on weekdays and 0.903 on holidays, 0.109 above the strongest external baseline (Greedy, 0.780). Second, energy (1210--1247~kJ, $\sim$3\% spread) and fairness discriminate weakly. All methods attain Jain's index no lower than 0.93 on weekdays and 0.88 on holidays; the geometry-based methods (Static, Greedy, INS-WOA) reach 0.999--1.000, since the per-UAV demand-proportional bandwidth split guarantees fairness structurally, and only the more stochastic RL methods incur measurable loss (JTORATC 0.930/0.888, EMORL-TCTO 0.946/0.978, HRL-TPRA 0.956/0.968, MADDPG 0.978/0.990, weekday/holiday). Hover power dominates the Zeng model \eqref{eq:zeng} and the battery state machine homogenizes flight patterns, so served ratio is the metric that separates policies; we report it as primary.

Greedy, GA-MATR, and TD-MPC cluster between 0.769 and 0.780 on weekdays: per-slot re-optimization and short-horizon forecasts capture most of the attainable service under regular commute demand. INS-WOA (0.744) sits between Static (0.736) and Greedy, consistent with its Voronoi pre-deployment inheriting K-means' sensitivity to demand shape. The hierarchical and multi-objective RL methods (EMORL-TCTO 0.614, HRL-TPRA 0.485) trail the table; Section~\ref{sec:lessons} reports the diagnostic chain behind these negative results.

\subsection{OOD Analysis}
\label{sec:ood}
Fig.~\ref{fig:ood} shows per-day served ratios across the Christmas--New Year OOD week. DSWM is the most stable method under shift: its holiday score (0.903) exceeds its weekday score. GA-MATR attains the highest holiday score among learning baselines (0.854, up from 0.770) and, together with TD-MPC (0.769$\to$0.805), is one of only two learning baselines surpassing Greedy (0.792) on holidays---graph attention with a fairness regularizer has genuine value under shift; even so, it trails DSWM by 0.119 on weekdays. JTORATC shows the reverse failure mode (holiday 0.495 below weekday 0.583), an OOD backfire consistent with BCD re-solves locking onto stale demand snapshots. Model-free RL baselines (PPO, SAC, MADDPG) also gain 0.07--0.09 on holidays, but from weekday scores 0.17--0.24 below DSWM's; DSWM improves by 0.014 from the highest base.

\subsection{Ablation}
\label{sec:ablation}
Fig.~\ref{fig:ablation} reports the two structural ablations and a counterfactual demand-scaling intervention $\mathrm{do}(s)$. Removing the decomposed service head and regressing $\SR_t$ directly from the latent state (\textit{w/o dec}) drops the weekday score from 0.889 to 0.769 ($-0.120$), returning the method to Greedy's level: the coverage gain comes from the differentiable service physics, not latent capacity. Removing the uncertainty penalty (\textit{w/o unc}) \emph{raises} the weekday score by $0.004$ (0.894 vs.\ 0.889), with the holiday score essentially unchanged (0.902 vs.\ 0.903): in our setting the term is insurance against overconfident imagined branches, not a gain source, consistent with offline model-based RL \cite{yu2020mopo} where pessimism matters most near the data boundary. The counterfactual $\mathrm{do}(s)$ probe in Fig.~\ref{fig:ablation}(b) scales the imagined demand by $s\in\{0.5, 2\}$ at planning time: DSWM stays near its $s{=}1$ reference under both scalings and degrades less than the w/o-dec variant, which again points to the differentiable service physics---not latent capacity---as the robustness source.

\subsection{Sample Efficiency and Scalability}
\label{sec:sample}
\begin{figure}[!t]
\centering
\includegraphics[width=\columnwidth]{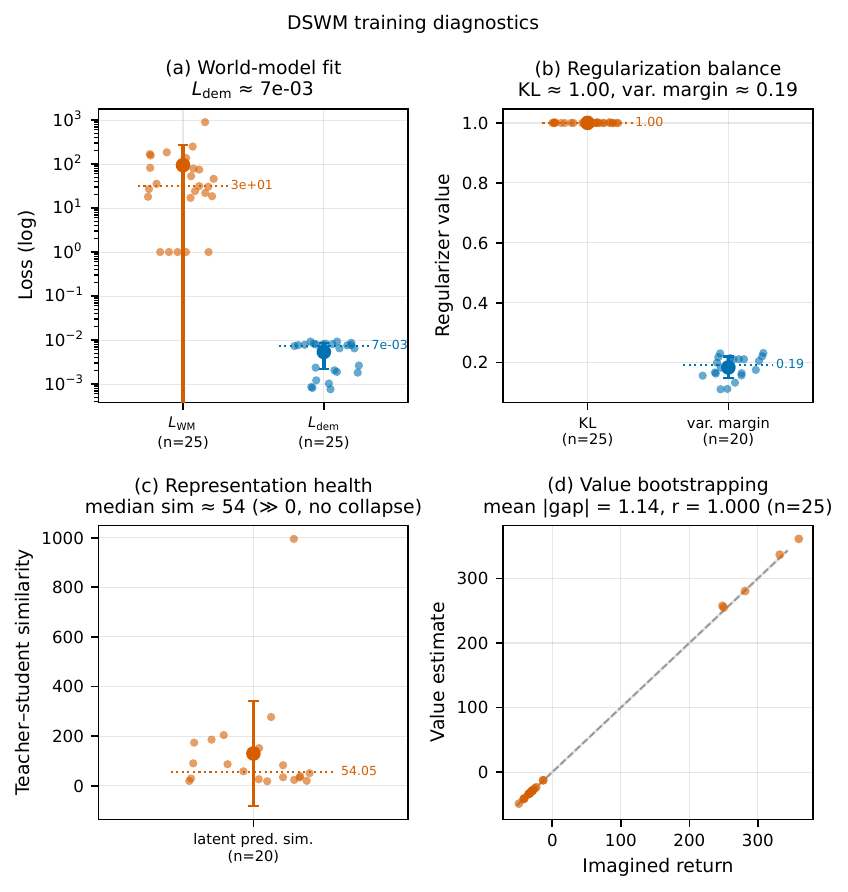}
\caption{Training diagnostics of DSWM: world-model loss, demand MSE, free-bits KL, and latent-prediction similarity terms. Probe curves use the test-day diagnostic protocol (probe-based, diagnostic only).}
\label{fig:train}
\end{figure}

Fig.~\ref{fig:sample} plots served ratio against environment steps (log scale) under an explicitly stated probe protocol: probe-based and diagnostic only (fixed test-day probe; fixed budget with no early stopping or model selection; probe numbers never enter Tables~\ref{tab:milan} or~\ref{tab:cross}). The x-axis counts environment interaction, including the 60{,}000 pre-collected transitions. DSWM's probe curve flattens early, which motivated the 30k-step gradient budget. Fig.~\ref{fig:scaling} varies the fleet size $K \in \{2, 8, 16\}$ and demand intensity: DSWM's margin over Greedy grows with $K$, since coordinating a 16-UAV fleet is exactly where myopic methods lose most. Fig.~\ref{fig:train} shows the training diagnostics (world-model loss, demand MSE, KL, and latent-prediction similarity terms); these use the same test-day probe protocol and are diagnostic only.

\subsection{World Model Quality}
\label{sec:quality}
\begin{figure}[!t]
\centering
\includegraphics[width=\columnwidth]{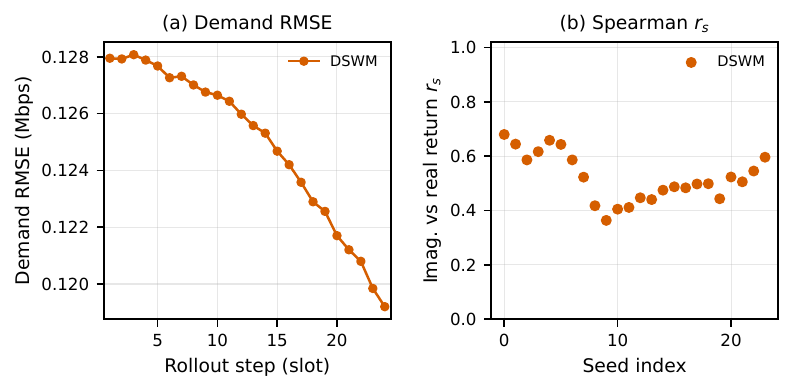}
\caption{World-model quality on the Milan test period. (a) Demand RMSE vs.\ imagination rollout step. (b) Spearman correlation $r_s$ between imagined and realized returns; each point is one seed--evaluation-day pair.}
\label{fig:quality}
\end{figure}

Fig.~\ref{fig:quality} evaluates the learned model along two axes: demand RMSE over the imagination horizon, and per-seed Spearman rank correlation $r_s$ between imagined and realized returns. Demand RMSE stays flat (0.128~Mbps at one step, 0.119~Mbps at 24 slots), and the imagined-return ranking correlates moderately with the realized one ($r_s=0.520\pm0.087$; each point in Fig.~\ref{fig:quality}(b) is one seed--evaluation-day pair): the model preserves \emph{where} demand concentrates far better than exact magnitudes---the quantity anchored planning consumes.

\begin{figure*}[!t]
\centering
\includegraphics[width=1.6\columnwidth]{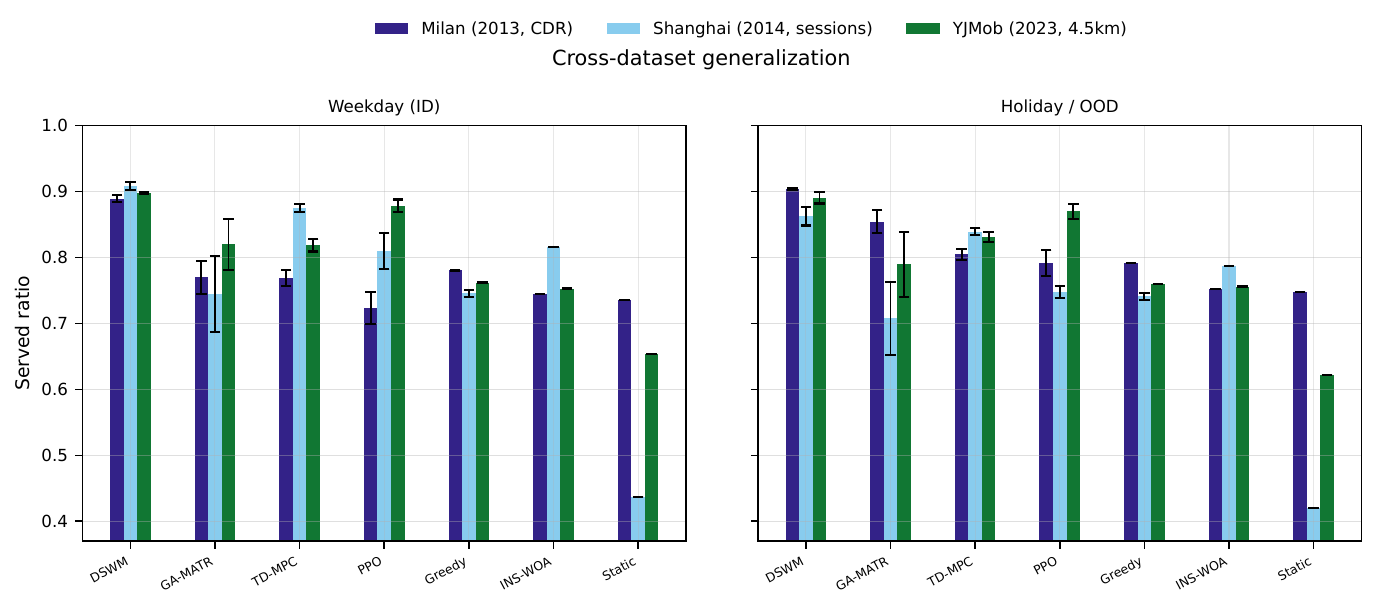}
\caption{Cross-dataset comparison (Milan (2013, CDR), Shanghai (2014, sessions), YJMob (2023, 4.5km); 7 methods; 5 seeds for Milan, 3 for Shanghai/YJMob). DSWM ranks first among non-ablated configurations on all three datasets under one configuration with no per-dataset tuning.}
\label{fig:cross}
\end{figure*}

\subsection{Cross-Dataset Generalization}
\label{sec:cross}
Table~\ref{tab:cross} and Fig.~\ref{fig:cross} extend the comparison to Shanghai and YJMob (2023, 4.5km)\footnote{The YJMob main version de-aggregates the native 30-min counts uniformly into 10-min slots within each cell (no information injected) and crops the densest $9\times9$ cells ($4.5\times4.5$~km at 500~m) to align the geographic scale and slot physics with Milan.} under the identical production configuration with no per-dataset tuning. DSWM ranks first among non-ablated configurations on all three datasets: 0.908 on Shanghai (+0.033 over TD-MPC), 0.898 on YJMob100K (+0.019 over PPO), and 0.889 on Milan (+0.109 over Greedy). The margin ordering is informative: Milan's demand is the most dispersed (top 5\% cells: 13\%) at a 4.7-km scale, where dynamic repositioning is most valuable and DSWM's margin largest; YJMob100K demand is more concentrated (22\%) and Shanghai's extremely sparse (85.3\% zero cells), where static or myopic coverage already captures much of the achievable service and margins shrink. Spatial scale and dispersion thus moderate the value of dynamic repositioning. Two failure cases support the same reading from the negative side: on Shanghai, Static collapses to 0.437 because K-means placement fails on 85\%-sparse demand; on YJMob100K, learning methods regain their normal ordering once slot physics and geographic scale match Milan's.

\begin{table}[!t]
\caption{Cross-Dataset Results: Weekday / Holiday Served Ratio (mean; 5 seeds for Milan, 3 seeds for Shanghai and YJMob; columns: Milan (2013, CDR), Shanghai (2014, sessions), YJMob (2023, 4.5~km)). Per-seed std where it affects adjacent rankings: DSWM $\pm$0.006 (Shanghai), $\pm$0.002 (YJMob); GA-MATR $\pm$0.058 (Shanghai), $\pm$0.039 (YJMob).}
\label{tab:cross}
\centering
\footnotesize
\setlength{\tabcolsep}{3pt}
\begin{tabular}{@{}lccc@{}}
\toprule
Method & Milan & Shanghai & YJMob \\
\midrule
\textbf{DSWM (ours)} & \textbf{0.889/0.903} & \textbf{0.908/0.862} & \textbf{0.898/0.891} \\
TD-MPC & 0.769/0.805 & 0.875/0.839 & 0.818/0.831 \\
PPO & 0.723/0.791 & 0.810/0.748 & 0.879/0.870 \\
GA-MATR & 0.770/0.854 & 0.745/0.708 & 0.820/0.790 \\
Greedy & 0.780/0.792 & 0.746/0.741 & 0.762/0.760 \\
INS-WOA & 0.744/0.752 & 0.816/0.788 & 0.753/0.756 \\
Static & 0.736/0.748 & 0.437/0.420 & 0.654/0.622 \\
\bottomrule
\end{tabular}
\end{table}

\subsection{Discussion: Lessons Learned}
\label{sec:lessons}
The development process produced findings that we report as lessons, including negative results.
\begin{enumerate}\itemsep=1pt \topsep=2pt
\item \textit{Per-agent embedding checks are mandatory for MARL baselines.} Our first GA-MATR re-implementation collapsed: two global graph-attention layers flattened per-UAV embeddings (cross-agent std $=0$), degenerating multi-agent learning into a single-agent policy; only a variance check and symmetry-breaking fixes recovered 0.770.
\item \textit{Hierarchical complexity can be a liability.} HRL-TPRA was kernel-launch bound (serial pointer decoding plus 16 per-UAV actor updates per macro-slot); even after a 5.7$\times$ speedup it ranked last among learning baselines (0.485).
\item \textit{Environment drift requires frozen same-day references.} Identical code rerun on a different day deviates by $\sim$$10^{-4}$ (BLAS/CPU variation).
\item \textit{Model-free methods are not always safe.} Static (K-means) collapses to 0.437 on 85\%-sparse Shanghai demand.
\item \textit{OOD strength exists outside world models.} GA-MATR's graph attention with a fairness term yields real robustness under shift (Section~\ref{sec:ood}).
\item \textit{Pre-deployment inherits clustering bias.} INS-WOA's Voronoi stage inherits K-means' sensitivity to demand shape, landing mid-table on all three datasets.
\end{enumerate}

\subsection{Qualitative Visualization}
\label{sec:qual}
\begin{figure*}[!t]
\centering
\includegraphics[width=0.24\textwidth]{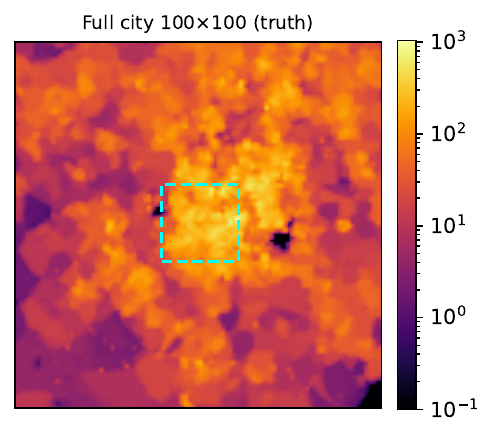}\hfill
\includegraphics[width=0.24\textwidth]{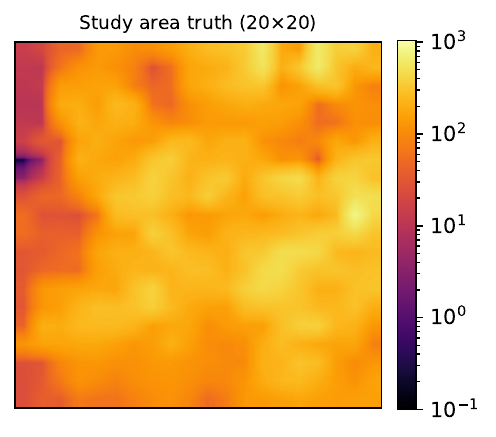}\hfill
\includegraphics[width=0.24\textwidth]{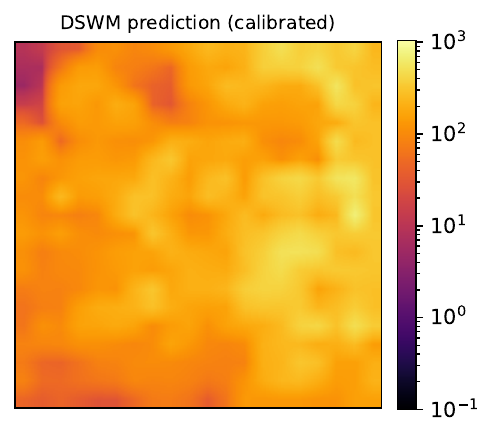}\hfill
\includegraphics[width=0.24\textwidth]{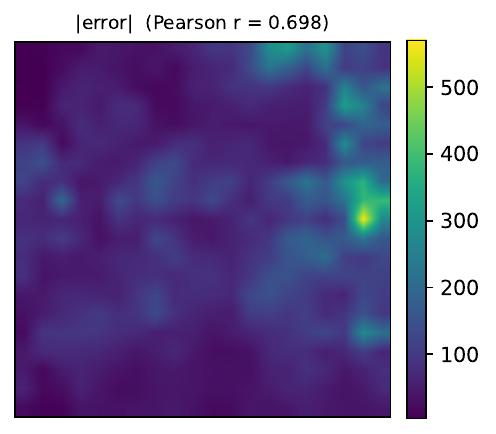}
\caption{World-model prediction visualization on Milan. From left to right: full-city context, ground-truth demand map, predicted demand map, and per-cell error. Predictions track the commute-driven hotspot migration; errors concentrate at cell boundaries during peak transitions.}
\label{fig:wmpred}
\end{figure*}

\begin{figure*}[!t]
\centering
\includegraphics[width=0.24\textwidth]{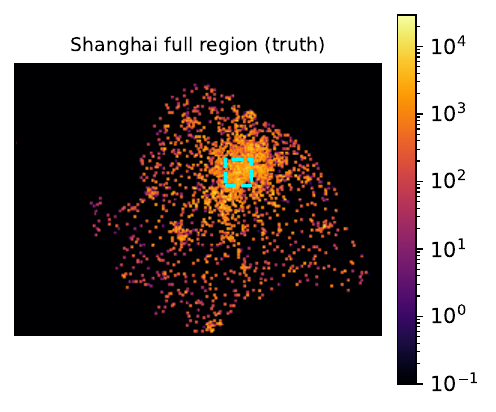}\hfill
\includegraphics[width=0.24\textwidth]{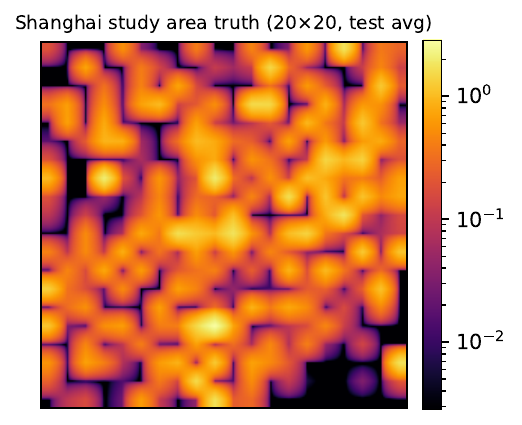}\hfill
\includegraphics[width=0.24\textwidth]{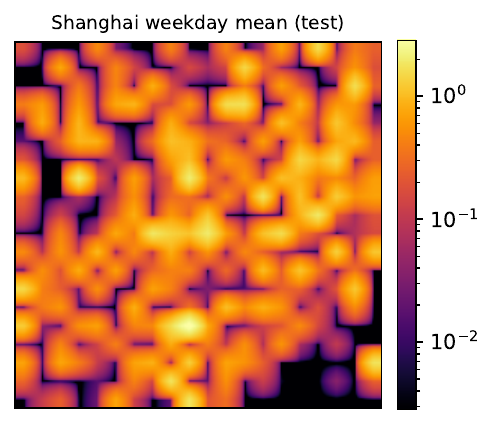}\hfill
\includegraphics[width=0.24\textwidth]{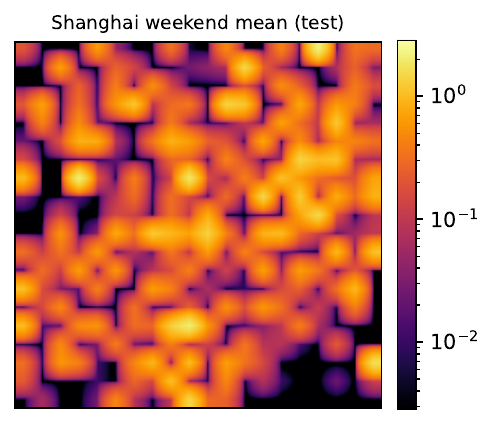}
\caption{Shanghai dataset visualization. From left to right: full-region demand, ground truth, weekday profile, and weekend (OOD) profile. The weekday--weekend shift is a spatial redistribution of sparse sessions (85.3\% zero cells), not a uniform intensity change.}
\label{fig:shvis}
\end{figure*}

\begin{figure*}[!t]
\centering
\includegraphics[width=0.24\textwidth]{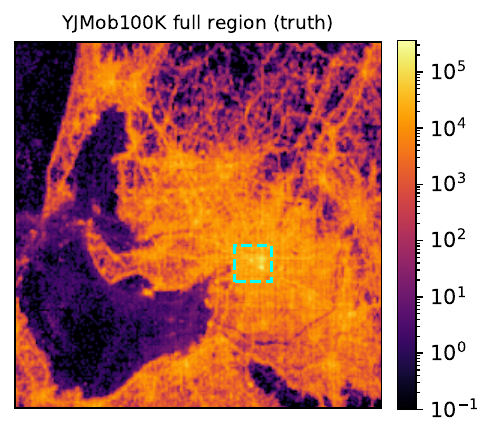}\hfill
\includegraphics[width=0.24\textwidth]{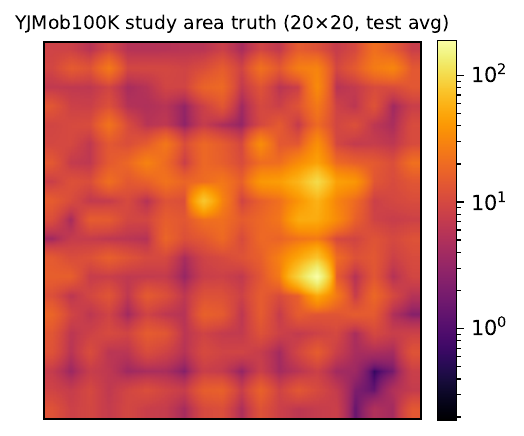}\hfill
\includegraphics[width=0.24\textwidth]{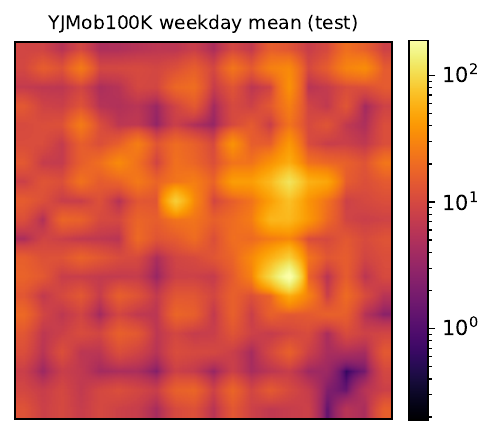}\hfill
\includegraphics[width=0.24\textwidth]{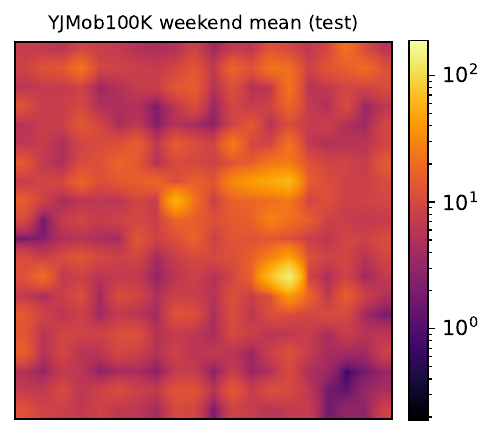}
\caption{YJMob100K dataset visualization. From left to right: full-region demand, ground truth, weekday profile, and pseudo-weekend (OOD) profile. Demand semantics are distinct-user presence counts (a 5\% sampled mobility proxy), as declared in Section~\ref{sec:data}; weekend labels are inferred from the autocorrelation-recovered week structure.}
\label{fig:yjvis}
\end{figure*}

Figs.~\ref{fig:wmpred}--\ref{fig:yjvis} give qualitative evidence. Fig.~\ref{fig:wmpred} compares the world model's predicted demand map against ground truth over the full Milan city grid, together with the error map: predictions track the commute-driven hotspot migration, and errors concentrate at cell boundaries during peak transitions. Figs.~\ref{fig:shvis} and~\ref{fig:yjvis} show full-region demand, ground truth, and weekday/weekend contrasts for Shanghai and YJMob100K; the weekday--weekend shift that defines each dataset's OOD split is visible as a spatial redistribution rather than a uniform intensity change, which is the regime where anchored planning helps most.

\subsection{Limitations}
\label{sec:limitations}
Two limitations qualify our claims. First, YJMob100K demand is a distinct-user presence proxy (5\% sampling), not traffic bytes, so cross-dataset conclusions about absolute served ratios should be read with the declared semantics. Second, energy and fairness discriminate weakly under hover-dominated power and battery constraints ($\sim$3\% energy spread, Jain 0.888--1.000), so our claims rest on served ratio. 

\section{Conclusion}
\label{sec:conclusion}
We formulated demand-driven UAV-BS fleet repositioning as a decision-time planning problem in a learned latent space and proposed DSWM, a world model of the demand-service dynamics built around three mutually reinforcing designs. The decomposed demand-service physics head makes the coverage mechanism differentiable inside the model, so the model is trained on the quantity the operator actually cares about rather than a latent proxy. The latent predictive regularization objective keeps the learned dynamics stable and collapse-free without reconstructing heavy-tailed demand maps. And observation-anchored CEM planning grounds every imagined rollout in the current observation instead of an open-loop forecast, which proves to be the decisive choice: across three real-world datasets, a broad lineup of learning, optimization, and classical baselines, and both in-distribution and holiday out-of-distribution weeks, DSWM consistently delivered the best demand coverage, including under demand regimes much sparser or more concentrated than the training distribution. The evidence supports the paper's central thesis: for this class of problems, the binding constraint on coverage is not how accurately demand is forecast, but how the current observation is exploited at decision time. The same design guideline carries over to other partially observed network control problems where the physical service model is known but the exogenous demand process is not, and more broadly to agentic network controllers that must perceive, remember, reason, and coordinate under uncertainty. Future work includes byte-accurate demand semantics on mobility datasets, heterogeneous fleets, and energy-aware altitude control.

\bibliographystyle{IEEEtran}
{\footnotesize\bibliography{references}}

\end{document}